%% file: main.tex
\documentclass[11pt,letterpaper]{article}

\usepackage[top=1in,bottom=1in,left=1in,right=1in]{geometry}
\usepackage{setspace}
\usepackage{fontspec}
\usepackage{microtype}
\usepackage{amsmath,amssymb}
\usepackage{graphicx}
\usepackage{booktabs}
\usepackage{tabularx}
\usepackage{multirow}
\usepackage{array}
\usepackage{caption}
\providecommand{\href}[2]{#2}
\usepackage{xurl}
\usepackage{natbib}
\usepackage{enumitem}
\usepackage{titlesec}
\usepackage{fancyhdr}
\usepackage{tikz}
\usetikzlibrary{shapes.geometric,arrows.meta,positioning,fit,backgrounds}
\usepackage{xcolor}
\usepackage{tcolorbox}
\tcbuselibrary{breakable}
\tcbset{rqbox/.style={colback=aaa-blue!7, colframe=aaa-blue, boxrule=0.5pt, arc=2pt, left=10pt, right=10pt, top=5pt, bottom=5pt}}
\tcbset{statusbox/.style={colback=highlight-gold!10, colframe=highlight-gold, boxrule=0.7pt, arc=2pt, left=10pt, right=10pt, top=6pt, bottom=6pt, breakable}}
\usepackage{ragged2e}
\usepackage{float}      

\bibpunct{(}{)}{;}{a}{ }{,}

\definecolor{aaa-blue}{RGB}{0,62,126}
\definecolor{light-gray}{RGB}{245,245,245}
\definecolor{highlight-gold}{RGB}{200,160,0}

\renewcommand{\thesection}{\arabic{section}}
\renewcommand{\thesubsection}{\thesection.\arabic{subsection}}
\renewcommand{\thesubsubsection}{\thesubsection.\arabic{subsubsection}}

\titleformat{\section}
  {\normalfont\large\bfseries}
  {\thesection.}{0.5em}{}
\titlespacing*{\section}{0pt}{18pt}{6pt}

\titleformat{\subsection}
  {\normalfont\normalsize\bfseries}
  {\thesubsection}{0.5em}{}
\titlespacing*{\subsection}{0pt}{12pt}{4pt}

\titleformat{\subsubsection}
  {\normalfont\normalsize\bfseries}
  {\thesubsubsection}{0.5em}{}
\titlespacing*{\subsubsection}{0pt}{10pt}{3pt}

\titleformat{\paragraph}[runin]
  {\normalfont\normalsize\bfseries}
  {}{0em}{}[.]
\titlespacing*{\paragraph}{1.5em}{8pt}{0.5em}

\begin{document}
\thispagestyle{plain}
\begin{center}
{\large\bfseries Human--AI Collaboration in Accounting Judgment:\\
A Proposed Cognitive Coding Scheme and a Rule-Based Adaptive Collaboration Protocol}
\vspace{12pt}

Yulu Huang\textsuperscript{1,2}\quad Lijun Xiao\textsuperscript{1}\quad
Gang Wang\textsuperscript{2,\ensuremath{\dagger}}

\vspace{6pt}
{\small\textsuperscript{1}Jiangxi University of Finance and Economics (JUFE), Nanchang, China\\
\textsuperscript{2}ByteDance}

\vspace{6pt}
{\small\textsuperscript{\ensuremath{\dagger}}Corresponding author: Gang Wang,
\texttt{wanggang@bytedance.com}\\
First-author email: \texttt{yuluhuang324@gmail.com}}

\vspace{6pt}
{\small Preprint prepared 22 November 2026}
\end{center}
\vspace{8pt}
\begin{abstract}
Large language models create an accounting information systems design problem:
how should evidence production, professional evaluation, and responsibility be
allocated between a model and an accountant? We propose a Cognitive
Decomposition for Accounting Judgment (CDAJ) coding scheme crossing four process
stages with information-intensive and evaluative demands. An Adaptive
Collaboration Protocol (ACP) connects allocation rules to an independent human
estimate, source-linked model output, structured verification, and a documented
human conclusion. Two separately returned coding forms cover 20 constructed
cases. A disclosed primary-only analysis retains unresolved auxiliary response
formats. Raw agreement is 19/20 for unitization, 20/20 for stage, and 18/20 for
demand. On 14 jointly single, codable cases, nominal Cohen's $\kappa$ is 1.000 for
stage and 0.752 for I/E/M demand profiles; only 12/20 cases are jointly resolved
to I or E. A public-disclosure diagnostic searches 1,406 pages for 18 numeric
reference cards from three banks' annual reports. Adding a value/period/unit
co-occurrence gate to BM25 with identical query information does not improve
designated-origin recovery at ranks one or three: both methods recover 13/18
origins at rank one and 18/18 within three ranks. A post-hoc audit distinguishes
page recovery from requested table-cell verification. Synthetic checks examine
event ordering and record integrity; Gaussian planning experiments examine a
future effectiveness evaluation. A separate source-response audit records adverse or unclear statuses in 16/36
returned records. These bounded findings concern coding repeatability and
implementation properties. They do not establish an optimal
allocation, a validated taxonomy, or improved professional judgment.
\end{abstract}

\clearpage
\section{Introduction}\label{sec:intro}
Professional accounting judgment combines evidence acquisition, analytical work,
and evaluation under uncertainty and accountability \citep{solomon1995,nelson2009}.
An LLM may assist with parts of this work without improving the resulting
professional conclusion. A correct extraction, a persuasive memorandum, and an
appropriately supported judgment are different system outputs. The accounting
information systems (AIS) design question is therefore how to organize the
interaction between evidence-producing technology and the professional who
reviews and owns the decision.

Prior AIS research connects technological adoption to the interaction of
organizations, regulation, and work practices \citep{austin2021,krieger2021}.
Research on AI in auditing also identifies ethical and accountability concerns
\citep{munoko2020}. The recent framework of \citet{stratopoulos2025} positions AI
within accounting research more broadly. Our focus is narrower: the interface
and record structure through which a professional evaluates a particular
AI-assisted accounting judgment. The professional's acceptance of AI advice is
not necessarily excessive. Accounting experiments document aversion to
AI-sourced advice, and the influence of user control depends on participant
characteristics \citep{comerford2022,comerford2024}. Thus neither blanket
acceptance nor blanket resistance provides a sound design objective.

We propose a Cognitive Decomposition for Accounting Judgment (CDAJ) coding scheme
and an Adaptive Collaboration Protocol (ACP). CDAJ crosses four stages of work
with two dominant cognitive demands. ACP converts the resulting allocation
rules into an inspectable sequence: record an independent estimate, expose
source-linked AI output, verify its evidence and reasoning, and document a
human conclusion. Here ``adaptive'' means that allocation and the availability
of iterative challenges vary with a task's requirements. It does not mean that
an implemented system learns an optimal policy or that a real-time adaptive
policy has been evaluated.

The proposal preserves a distinction between \emph{normative constraints}
and \emph{performance predictions}. Retaining a responsible human signatory is
a design constraint; whether an independent estimate or a checkpoint improves
judgment is an empirical question. This distinction prevents professional
accountability from being mistaken for evidence that a human-led configuration
is always more accurate.

\subsection{Research questions and contributions}
The paper asks three questions. RQ1: how can process stage and cognitive demand
organize accounting-specific allocation rules? RQ2: how can these rules be
implemented as inspectable interface requirements and decision records? RQ3:
what evidence would distinguish useful protocol increments from unnecessary
burden? We answer the first two conceptually and provide a public-source ambiguity
diagnostic, computational checks, and a human-evaluation agenda for the third.

The contribution has three parts. First, CDAJ connects an eight-cell allocation
matrix to 25 illustrative accounting tasks, identifying where information
production can be delegated while evaluation and sign-off remain explicit.
Second, ACP links each safeguard to observable interface states and record
fields. Its accounting-specific content lies in preserving source provenance,
challenged assumptions, admissible judgment ranges, and ownership through the
whole working-paper chain, rather than merely adding an explanation to an AI
recommendation. Third, an executed public-disclosure ambiguity diagnostic exposes the
difference between locating a page and verifying an accounting claim;
reproducible calculations and synthetic checks examine additional operating
properties. A separate exercise examines primary coding repeatability for 20
constructed cases using two returned response forms, with a disclosed limited
execution while auxiliary response formats remain unresolved. Neither the
computational checks nor coding agreement shows that accountants reason better
under ACP.

This paper proposes an AIS coding scheme and a record and workflow specification
with bounded technical evaluation and a limited coding exercise, informed by design-science principles \citep{hevner2004}. It is not
a completed evaluation of artifact utility in professional practice, a
registered report, or a preregistered human study. The formal state and record specifications are inspectable design artifacts.
The three cases are design illustrations; complete participant-facing evidence files and adjudicated
scoring keys remain future work. The paper retains an explicit evaluation
agenda because an auditable process can still produce poor judgments, and its
costs may exceed its benefits.

\section{Theoretical grounding}\label{sec:literature}
\subsection{Accounting judgment and task structure}
Accounting judgment research shows why the task, evidence, decision maker, and
review environment must be considered together \citep{libby1976,bonner1999,
solomon1995}. Professional skepticism involves both skeptical judgment and
action, so a recorded concern is not equivalent to a consequential verification
step \citep{nelson2009}. Early auditing research also examines anchoring and
adjustment in probabilistic inference \citep{joyce_biddle1981}. These findings
motivate a workflow that records an initial position and subsequent evidence
use, but they do not establish that precommitment is always helpful: an initial
human error can itself become an anchor.

CDAJ distinguishes information-intensive work from evaluative work as a
\emph{dominant-demand coding}, not as mutually exclusive mental faculties.
Extracting a covenant may be mechanically checkable; determining whether an
uncertain refinancing plan changes a going-concern conclusion requires
contextual evaluation. A whole engagement contains both. Accounting domain,
process stage, and cognitive demand should therefore be coded separately.
An expected-credit-loss case is not intrinsically information-intensive simply
because it contains numerical computation.

\subsection{Allocation, complementarity, and reliance}
Levels of automation and mixed-initiative interaction provide the general
basis for allocating functions between people and systems
\citep{parasuraman2000,horvitz1999}. Trust should be calibrated to system
capability and context \citep{lee2004}; reliance is not a quality measure in
itself. Human--AI complementarity requires examining the combined system, not
inferring team performance from either agent alone \citep{hemmer2025}.
Field evidence outside accounting finds uneven effects of AI assistance across
knowledge tasks \citep{dellacqua2026}, reinforcing the need for task-contingent
evaluation without establishing the same effects for accountants.

Accounting evidence adds an important countermechanism. \citet{comerford2022}
study auditors' reliance on AI in complex estimates, while
\citet{comerford2024} examine how providing input affects reliance, with the
control effect contingent on external locus of control. These studies make
algorithm aversion a plausible alternative to automation bias. A checkpoint
could correct excessive deference, but could also encourage unnecessary
rejection of sound advice. The relevant outcome is the quality and defensibility
of the resulting judgment, with reliance behavior interpreted in that light.

\subsection{Verification and cognitive forcing}
Automation bias concerns inappropriate dependence on automated assistance
\citep{parasuraman2010,goddard2012}. Explanations can alter reliance without
necessarily improving human--AI team performance \citep{bansal2021}.
\citet{bucinca2021} investigate cognitive forcing as a way to reduce
overreliance, including its costs. ACP draws on that design logic but does not
assume its effectiveness transfers unchanged to professional accounting.
Mandatory verification can consume scarce time, become ceremonial, or strengthen
a user's attachment to an initially mistaken view.

Professional review requires inspectable evidence as well as an intelligible
rationale. A fluent rationale may contain unsupported claims; token likelihood
is not a probability that an accounting conclusion is correct. ACP therefore
separates source verification, uncertainty disclosure, and any numerical
calibration claim. The IAASB's support material on technology overreliance
provides practice context \citep{iaasb2021}; it does not certify this protocol's
compliance with auditing standards.

\subsection{From general principles to an accounting design}
The proposed extension is a traceable chain linking four design elements:
allocation by stage and demand; independent human evidence assessment;
claim-level source comparison; and a signed rationale explaining accepted and
rejected advice. Generic human--AI principles already motivate human control
and explanation \citep{amershi2019}. CDAJ/ACP specifies how these principles
interact with range-valued accounting estimates, competing standards-based
interpretations, and engagement review. The inventory and cases demonstrate
this mapping, not the exhaustiveness or superiority of the coding scheme.
Alternative decompositions and stronger existing workflows could perform as
well or better; the evaluation must retain that possibility.
\section{Cognitive coding scheme for accounting judgment}
\label{sec:innovation1}

\subsection{Motivation: From AI Capability to Task Allocation}

A capability-first approach to LLM deployment in accounting asks,
``What can the LLM do well?'' and routes tasks accordingly.
This capability-first perspective is incomplete because it ignores
the second-order question: \emph{what should the human continue to
do}?
Capability benchmarking answers whether an LLM can draft a
memorandum or estimate an allowance; it does not answer whether
allowing it to do so unconditionally improves the firm's audit
quality, regulatory posture, or stakeholder trust.

This section develops a complementary perspective: a
\textbf{judgment-first} decomposition that classifies accounting
decision tasks along two dimensions---process stage and cognitive
demand---and produces principled predictions about task allocation
between human and AI agents.

\subsection{Process-Stage Decomposition}

Adapting \citet{solomon1995} and \citet{parasuraman2000}, we
decompose a canonical accounting judgment into four sequential
process stages:

\begin{enumerate}[leftmargin=2em, label=\textbf{\arabic*.}]
  \item \textbf{Data Acquisition} --- locating, extracting, and
    normalizing evidence from engagement files, regulatory
    databases, prior working papers, and external sources.

  \item \textbf{Analytical Synthesis} --- computing ratios,
    trends, comparisons to benchmarks, statistical
    anomaly detection, and the construction of coherent
    evidence packages.

  \item \textbf{Evaluative Judgment} --- forming a professional
    opinion (e.g., the going-concern assumption is appropriate;
    the allowance is sufficient; the Level-3 estimate is
    reasonable) under conditions of uncertainty and materiality.

  \item \textbf{Professional Communication} --- documenting
    the judgment in a memorandum, footnote, or audit report;
    explaining the basis for the conclusion to stakeholders;
    and accepting personal accountability for the opinion.
\end{enumerate}

The four stages differ in the type of cognitive work they demand,
the verifiability of the output, and the accountability attached
to the outcome.
This motivates a second decomposition along the axis of cognitive
demand.

\subsection{Cognitive-Demand Decomposition}

We classify each stage along a single dimension: whether the work
is predominantly \emph{information-intensive} (requires recalling
or computing over a large, structured body of evidence) or
\emph{evaluative} (requires weighing competing considerations,
exercising professional skepticism, and accepting accountability
for a judgment under uncertainty).

Information-intensive work rewards consistency, completeness, and
computational accuracy.
Evaluative work rewards contextual awareness, ethical reasoning,
and the willingness to challenge precedent.

LLMs can assist with retrieval, extraction, and drafting, but performance must
be assessed against task-specific references. Deterministic calculations should
be executed in validated computational tools rather than trusted to fluent text
generation. Evaluative tasks add unresolved interpretation, professional
skepticism, and responsibility. These properties motivate conditional allocation
rules; they do not establish a universal capability advantage for either agent.

\subsection{The Allocation Matrix}

Crossing the four process stages with the two cognitive demands
yields the $4 \times 2$ matrix of Table~\ref{tab:matrix}, which
maps each cell to a proposed allocation hypothesis for evaluation.

\begin{table}[H]
  \singlespacing
  \centering
  \caption{CDAJ Allocation Matrix: Proposed Allocation Hypotheses by
    Process Stage and Cognitive Demand.}
  \label{tab:matrix}
  \begin{tabular}{@{}>{\RaggedRight\arraybackslash}p{2.8cm}>{\RaggedRight\arraybackslash}p{4.3cm}>{\RaggedRight\arraybackslash}p{4.3cm}@{}}
    \toprule
    \textbf{Process Stage} & \textbf{Information-Intensive Tasks}
      & \textbf{Evaluative Tasks} \\
    \midrule
    Data Acquisition
      & LLM assists extraction; human checks coverage
      & Human leads (curation requires context) \\
    \addlinespace
    Analytical Synthesis
      & LLM leads with human verification
      & Iterative: LLM produces candidates; human selects and defends \\
    \addlinespace
    Evaluative Judgment
      & Iterative: LLM produces structured analysis; human
        issues opinion under skepticism
      & Human leads; LLM as adversarial reviewer \\
    \addlinespace
    Professional Communication
      & LLM leads drafting; human leads tone and sign-off
      & Human leads; LLM supports formatting and citation \\
    \bottomrule
  \end{tabular}
  \vspace{4pt}\\
  \footnotesize\textit{Note}: ``LLM leads'' denotes primary
  generation by the AI agent with human review; ``iterative''
  denotes alternating contributions with explicit checkpoints;
  ``Human leads'' denotes human authorship with LLM assistance
  available on demand. These are untested allocation hypotheses, not an
  operational default or authorization to delegate. Coding agreement does not
  establish their relative quality, efficiency, or safety.
\end{table}

The matrix generates several non-obvious predictions.
First, even in stages where the LLM assists extraction
(Data Acquisition), the human retains a verification role: the
LLM's retrieval coverage and factual accuracy are uncertain, and
omissions in evidence collection propagate through the entire
downstream analysis.
Second, the matrix explicitly rules out ``LLM-only'' deployments
in evaluative-judgment and professional-communication stages
for tasks classified as evaluative: these stages are exactly
where professional accountability attaches.
Third, the matrix identifies the Analytic Synthesis stage as
the most natural site for collaboration: both information-intensive
and evaluative elements coexist, and the protocol must support
both modes.

\subsection{Coding rules and boundary conditions}\label{sec:coding}
CDAJ is a proposed coding heuristic, not an empirically validated taxonomy.
It classifies a \emph{work unit}, defined as an action with a specified input,
a distinct output, and an acceptance criterion. It does not classify a whole
engagement by its job title. The coding procedure has four steps.
\begin{enumerate}[leftmargin=2em]
\item Segment work whenever its output or acceptance criterion changes.
Extracting a loan balance and choosing an overlay are separate units even when
the same person performs both.
\item Assign the stage by the unit's immediate output: a located or normalized
source is Data Acquisition; a derived comparison or scenario is Analytical
Synthesis; a selected conclusion is Evaluative Judgment; a representation of
that conclusion for review or disclosure is Professional Communication.
\item Code demand as information-intensive when the applicable criteria and
assumptions are already fixed and the output can be checked by tracing or
recomputation without resolving a disputed interpretation. Code it as
evaluative when acceptance requires choosing among defensible assumptions,
assessing source credibility, or weighing competing interpretations. The stage
names refer to outputs; the demand labels refer to how acceptance is decided.
Thus applying an already approved covenant threshold is a judgment-stage
information-intensive unit, whereas choosing which management assumption is
credible is evaluative even during data acquisition.
\item When both demands remain material, split the unit if its outputs can be
separated; otherwise retain both candidate codings, route the unresolved choice
to a human-led checkpoint, and document the ambiguity. Do not force a majority
label merely to populate the matrix.
\end{enumerate}
We collected two separate coder returns for 20 constructed case descriptions
using these rules. Section~\ref{sec:human-coding} distinguishes the available
primary coding analysis from unresolved auxiliary response formats. Independent
labels are retained; no reconciled labels are used to calculate agreement.
Appendix~\ref{app:matrix} supplies 25 author-selected work-unit examples with
proposed dominant codings, not a validated universe or reliability estimate.
The following decomposition makes the coding inspectable in each main domain.

\begin{table}[H]\centering\singlespacing\small
\caption{Illustrative work-unit decomposition under the explicit coding rules.}
\label{tab:units}
\begin{tabularx}{\textwidth}{@{}>{\RaggedRight\arraybackslash}p{2.0cm}>{\RaggedRight\arraybackslash}p{2.7cm}XX@{}}
\toprule
Case & Stage & Information-intensive unit & Evaluative unit\\\midrule
Going concern & Acquisition & Extract dated covenant terms & Assess credibility of financing evidence\\
 & Synthesis & Recompute a specified coverage ratio & Choose defensible forecast stress assumptions\\
 & Judgment & Apply an approved threshold to checked inputs & Weigh contrary evidence in the conclusion\\
 & Communication & Assemble source cross-references & Defend the conclusion and its caveats\\\addlinespace
IFRS~9 ECL & Acquisition & Extract loan balances and dates & Assess completeness of forward-looking evidence\\
 & Synthesis & Execute an approved scenario calculation & Select scenario weights or overlay assumptions\\
 & Judgment & Check an estimate against approved constraints & Decide whether qualitative adjustment is justified\\
 & Communication & Reconcile disclosure figures & Explain significant judgment and uncertainty\\\addlinespace
Level-3 value & Acquisition & Extract transaction terms & Assess comparability of a candidate transaction\\
 & Synthesis & Execute a specified DCF calculation & Choose defensible unobservable inputs\\
 & Judgment & Compare value with an approved admissible range & Select and defend a value among alternatives\\
 & Communication & Assemble input and sensitivity schedules & Explain valuation uncertainty and limitations\\
\bottomrule
\end{tabularx}
\end{table}
The matrix supplies candidate allocations for testing on these coded units.
Evidence access and independently checked component capability are proposed
necessary conditions, not evidence that the allocation is beneficial. It neither substitutes
for professional standards nor shows that an LLM is superior in an
information-intensive cell. Human capability, model quality, time pressure,
error consequences, and existing review controls may justify another
allocation. The distinctive design commitment is that changing the generation
lead cannot erase the source-to-judgment record or transfer final responsibility
to the generator.

\subsection{Worked boundary cases and delegation gates}\label{sec:delegation}
Three boundary cases clarify what the coding rules commit to. These are worked
conceptual examples, not observations from a coding-reliability study.

\paragraph{Credit-loss calculation versus overlay selection}
Suppose the input is a loan portfolio with approved stages, scenarios, weights,
and model parameters. Producing the allowance schedule by executing that fixed
specification is Analytical Synthesis / information-intensive; acceptance is
independent recomputation against the approved inputs. If the requested output
also includes selecting an overlay because management disputes the macroeconomic
forecast, that is a distinct evaluative work unit. Its acceptance requires
reasons for selecting one assumption over another. Split the schedule execution
from overlay selection and link the records. If the effects cannot be separated,
retain the mixed coding and prohibit unconditional LLM-led completion.

\paragraph{Financing evidence versus a going-concern conclusion}
Extracting a signed facility's date, amount, and conditions is Data Acquisition /
information-intensive when the source identity is established. Deciding whether
a management email is credible evidence of imminent refinancing is Data
Acquisition / evaluative: the immediate output is accepted evidence, not yet the
final opinion. Weighing that evidence with cash-flow shortfalls to reach a
conclusion is Evaluative Judgment / evaluative. This boundary prevents a single
``going concern'' label from conflating retrieval, credibility assessment, and
professional conclusion.

\paragraph{Valuation arithmetic versus input admissibility}
Executing DCF arithmetic under fixed discount rates and cash flows is Analytical
Synthesis / information-intensive. Choosing whether a transaction is comparable
or an unobservable input is defensible is evaluative. Once an authorized review
has fixed an admissible range, checking whether a computed value falls within
it is Evaluative Judgment / information-intensive under the output-based stage
rule. Selecting or defending that range is not: it remains evaluative.
A unit that asks the model both to set the range and certify its own value
therefore fails the fixed-acceptance-criterion condition and must be split or
kept human-led. The scheme can be challenged by identifying work units for
which these stated boundaries give incompatible classifications.

The matrix's LLM-leading cells nominate allocations to evaluate; they do not
grant permission to delegate. We propose five necessary preconditions to
document before evaluating such an allocation:
(i) the acceptance criteria and material assumptions are fixed;
(ii) authorized source access and a material-evidence manifest are available;
(iii) local capability checks pass for the particular component and version;
(iv) material source links and calculations can be independently checked; and
(v) an accountable reviewer can resolve exceptions. An unknown gate counts as
unmet, not as permission. In evaluative or unresolved mixed units, human-led
resolution remains required regardless of the model's fluency.

For deterministic arithmetic, gate (iii) requires agreement with independent
recomputation under the same inputs, units, and declared numerical tolerance;
a discrepancy blocks model-led completion. For retrieval, the owner must define
material reference items and an acceptable coverage/error threshold on a relevant
local validation set before testing. Every designated critical omission must be
resolved; passing a finite set does not establish population recall. There is
no universal numerical threshold asserted by this paper and no claim that a
particular LLM has passed these gates. If a threshold or relevant validation set
is absent, delegation remains unapproved. A failed capability/version check
routes the work to reviewed deterministic tools or human-led processing; missing
material evidence triggers escalation or withholding a conclusion, rather than
permitting either agent to invent it. The gate result, tested version, exceptions,
and reason for the chosen allocation belong in the judgment record. Satisfying
these proposed preconditions is not sufficient evidence of safe or beneficial
delegation: comparative evaluation and applicable professional authorization
remain separate requirements.

\subsection{Significance for AIS Design}

The proposed CDAJ scheme organizes three design principles for the design
of AI-assisted accounting information systems.
First, \emph{design by process stage}, not by job title: the
allocation of work between human and AI should be specified at
the level of each stage of each recurring task, not at the
level of ``the auditor'' or ``the analyst.''
Second, \emph{preserve accountability boundaries}: under the proposed accountability constraint, evaluative
judgments and professional sign-off remain human-led, because accountability cannot be
delegated to a system that does not bear professional liability.
Third, \emph{instrument collaboration, not replacement}: the
metric of success is the quality of the team's output, not the
fraction of work performed by AI.

\section{Adaptive Collaboration Protocol}
\label{sec:innovation2}

\subsection{Design Goals}

The CDAJ allocation matrix of Section~\ref{sec:innovation1} specifies
\emph{who should do what}, but not \emph{how} the collaboration should
proceed.
This section operationalizes the matrix as the \textbf{Adaptive
Collaboration Protocol} (ACP), a verifiable workflow that embeds
three proposed safeguards against inappropriate reliance, whose accuracy
and time costs require evaluation.

The ACP has three design goals:
\begin{enumerate}[leftmargin=2em]
  \item \textbf{Preserve professional accountability} by ensuring
    that every documented conclusion is owned by a credentialed human
    who can defend the position under professional review.

  \item \textbf{Defend against automation bias} by inserting interface
    features intended to interrupt premature convergence on AI suggestions.

  \item \textbf{Support verifiability} by requiring that every AI
    material claim be accompanied by source references and explicit
    uncertainty disclosure.
\end{enumerate}

\subsection{Protocol Architecture}

ACP specifies a four-stage workflow for material judgment work units.
The event harness in Section~\ref{sec:evaluation} implements selected
event-order checks, separately from the record validators and public-source
experiment; none implements a participant-facing platform.

\subsubsection{Stage A: Independent Human Estimate}

Before viewing any LLM output, the human professional records an
\emph{independent estimate} of the conclusion, the key supporting
evidence considered, and the confidence level (0--100 percent).
The purpose is to preserve an observable pre-advice position and encourage
independent evidence consideration \citep{bucinca2021,joyce_biddle1981}.
Whether it reduces harmful reliance, or instead reinforces an incorrect human
anchor, is an open behavioral question. An initial position is not ground truth.

The proposed interface would hide the LLM panel until the independent
estimate has been recorded; visual markers indicate the protocol
status (``estimate pending'' vs.\ ``estimate recorded'').

\subsubsection{Stage B: Source-linked output and uncertainty disclosure}
The model output contains a proposed conclusion, a claim-by-claim rationale,
source locations within the evidence file, and an uncertainty panel identifying
missing information, unresolved assumptions, and conditions that would change
the recommendation. The human must be able to open the cited source and check
whether it supports the claim. A reference supplied by a model is not itself
verified evidence; broken or non-supporting references are flagged for review.

Numerical confidence is optional and must not be labeled calibrated without
an independently evaluated prediction target, held-out reference cases,
calibration procedure, and uncertainty estimates for calibration performance.
Model-generated self-confidence and token probabilities do not satisfy that
requirement. The illustrative workflow and proposed first human evaluation
therefore use qualitative uncertainty disclosure, not an unvalidated percentage.
Evidence for calibration could be developed later on a sufficiently large,
relevant case set; three frozen vignette outputs cannot supply it.

\subsubsection{Stage C: Structured Verification Checkpoint}

The human professional reviews the LLM output against three
verification questions, recorded as free-text responses plus
Likert-scale ratings:
\begin{enumerate}[leftmargin=2em, label=\textbf{Q\arabic*.}]
  \item \emph{Agreement}: Do you concur with the LLM's
    recommendation?
    (Likert: 1 = strong disagreement, 7 = strong agreement)

  \item \emph{Evidence addition}: What evidence did you consult
    that the LLM did not, or vice versa?

  \item \emph{Reasoning divergence}: Where, if anywhere, does your
    reasoning diverge from the LLM's stated rationale?
\end{enumerate}
The questions are intended to elicit agreement, omitted evidence, and
reasoned disagreement. These are proposed interface diagnostics, not a
validated taxonomy of automation-bias mechanisms.

\subsubsection{Stage D: Documented Conclusion and Audit Trail}

The final output is a structured record containing:
(i)~the human's independent estimate from Stage A;
(ii)~the source-linked model output and uncertainty disclosure from Stage B;
(iii)~the verification responses from Stage C; and
(iv)~a final conclusion that may or may not match the LLM's
recommendation, signed by the credentialed human professional.

The record supports review of which evidence and assistance informed the
conclusion. It does not by itself establish compliance with any auditing
standard, the sufficiency of audit evidence, or the truthfulness of logged
content. Access controls, retention rules, signatures, source authorization,
and tamper-evident storage require separate implementation and assurance.
Differences between initial, model, and final answers describe revision and
reliance; their value cannot be determined without an independent quality
criterion.

\subsection{Formal event and record specification}\label{sec:formal}
ACP consists of two separable artifacts: an event-order acceptor and a judgment
record schema. The event acceptor determines whether the prescribed sequence
was followed; the record schema determines whether the required objects and
references are present. Neither determines whether the evidence is true or the
judgment professionally sound. This separation gives system designers explicit
failure conditions without equating procedural completeness with quality.

\begin{table}[H]\centering\singlespacing\small
\caption{Reference event-state specification for one work unit.}
\label{tab:states}
\begin{tabularx}{\textwidth}{@{}>{\RaggedRight\arraybackslash}p{2.6cm}>{\RaggedRight\arraybackslash}p{2.1cm}XX@{}}
\toprule
Current state & Event & Guard & Next state\\\midrule
Start & Independent & Initial-confidence field in range; first event & A: initial recorded\\
A & Payload & C2--C4 only; version pin and frozen-content hash match & B: advice exposed\\
B & Verification & C3--C4 only; agreement in range and required text present & V: checkpoint recorded\\
V & Challenge & C4 only; no pending challenge; nonempty text & Q: response pending\\
Q & Response & Matching challenge ID; version and response-content hash valid & V: checkpoint recorded\\
A / B / V & Final & C1 from A; C2 from B; C3/C4 from V; no pending exchange & D: finalized\\
D & Any event & No further transition allowed & Reject trace\\
\bottomrule
\end{tabularx}
\end{table}
All unlisted transitions, unknown events, and missing finalization are rejected.
The implemented event checker encodes Q as a pending-response flag while
retaining V; the table is the equivalent explicit-state representation.
A hash checks consistency with an expected payload, not authenticity of its
source. The reference implementation assumes strict timestamp ordering;
a deployment needs a unique event sequence even when clocks tie. Rejection
means that the trace does not satisfy the specification. Operational repair,
exception escalation, access controls, and preservation of rejected events
require platform implementation and are not implemented by the checker.

\begin{table}[H]\centering\singlespacing\small
\caption{Core judgment-record objects and accounting-specific integrity rules.}
\label{tab:record}
\begin{tabularx}{\textwidth}{@{}>{\RaggedRight\arraybackslash}p{3.0cm}XX@{}}
\toprule
Object & Required contents & Inspectable relation or failure\\\midrule
Context & Record ID, framework, measurement date, synthetic marker for fixtures & Do not interpret a decision outside its stated framework/date\\
Evidence registry & Source ID, version, content hash, source locator & Every cited source ID resolves; changed bytes invalidate the stored hash\\
Claims and assumptions & Claim ID, statement, source/assumption IDs, declared support status & A conclusion's inputs remain traceable; unsupported/contested claims require an exception\\
Initial estimate & Value, cited evidence, rationale, confidence & The pre-advice position is preserved separately from model advice\\
Verification & Agreement, added evidence, reasoning divergence, exception list & Each material exception identifies the claim being challenged\\
Exchanges & Exchange ID, challenge and response text & Optional iterative dialogue is retained rather than replacing initial advice\\
Final record & Value, conclusion, rationale, exception dispositions, sign-off & Every listed exception receives exactly one documented disposition before finalization\\
\bottomrule
\end{tabularx}
\end{table}
The schema distinguishes a claim's \emph{declared} support status from verified
truth. For each contested or unsupported claim, verification must record an
exception. Finalization requires a reasoned disposition: addressed, accepted
with reason, or decision withheld. Acceptance with reason does not turn weak
evidence into sufficient evidence; the final reviewer remains responsible for
the disposition. The implementation checks the existence and references of
these fields, not that an explanation is adequate or a signatory is authentic.
In production, authorized identity and protected storage must bind the recorded
sign-off to a person; the test implementation uses a fabricated signer ID.

The extended record contract makes accounting context typed rather than a
single descriptive label. It requires a framework, accounting assertion,
measurement date, explicit three-letter currency, unit, and guidance references.
A guidance reference either identifies an authority and paragraph or records
that applicability is unresolved with a reason. The admissible-range object
records established, not-established, or not-applicable status and a boundary
rationale; only an established range has finite ordered bounds. Thus a missing
professional basis is represented as missing knowledge rather than populated
with an invented standard citation or numerical range.

For a tabular claim, the extended contract requires the requested reference
binding and a typed source binding: source ID, document/version, one-based page,
table, row, column, accounting measure, period, currency, unit, printed literal,
and normalized value. Each binding must match a separately supplied source
catalog entry and resolve to the cited evidence registry. The checker preserves
currency as an explicit field rather than inferring it from a stripped symbol.
It checks numeric normalization and occurrence of the value and period in the
bound source text. This adds an inspectable relation between a requested
accounting measure and its source location; it does not infer that relation
from free text or reconstruct the table automatically.

The catalog is a trusted \emph{input assumption}, not a truth created by the
validator. It must be established and reviewed independently of the advice
being checked in an operational deployment. Matching a catalog protects an
already specified binding against inconsistent substitution; it cannot certify
an incorrect catalog or a claim whose free text contradicts its structured
fields. The executed extension below explicitly includes a case exposing this
limitation. Claim entailment, standards applicability, source authenticity,
and professional judgment remain distinct from typed-record consistency.

An explicitly fabricated loan-portfolio record demonstrates the distinction.
Given an exposure of 1,000,000 currency units and assumed PD of 0.03 and LGD of
0.40, the arithmetic product is 12,000. The record labels this a calculation
illustration, not a complete IFRS~9 allowance. It records an exception that
staging, discounting, and forward-looking scenario adequacy have not been
established. The final disposition withholds a professional allowance
conclusion while retaining the arithmetic value and the reason for withholding.
All source text, the initial value, and the nominal challenge/response in this
fixture are fabricated test inputs, not observations, expert judgments, or
LLM outputs. The example shows a concrete accounting use of the schema:
computational completion and professional conclusion are distinct states of
knowledge, and the record can preserve the former without claiming the latter.

The operational adaptation is therefore \emph{rule-based}: use the task coding
to determine the generation lead, require a human-led resolution when
interpretation remains unsettled, and preserve the same evidence/exception
chain through any allowed refinement. ACP does not optimize that routing
policy from observed outcomes. Its proposed contribution is an explicit link
between work-unit allocation and inspectable accounting judgment records,
which general advice explanations alone do not specify.

\subsection{Three Instantiations: Audit, Reporting, and Valuation}

We illustrate the ACP in three accounting scenarios spanning the
audit, financial-reporting, and valuation domains.

\subsubsection{Going-Concern Assessment (Audit Domain)}

In an illustrative US public-company going-concern assessment, using the
applicable PCAOB framework, an audit senior considers whether substantial doubt about the client's ability to continue as a going
concern exists.
The ACP specifies:
\begin{itemize}[leftmargin=2em]
  \item \textbf{Data Acquisition}: LLM proposes extracted
    debt-covenant-compliance evidence, recent management
    forecasts, and regulatory correspondence from the engagement
    file. Human curates for relevance.
  \item \textbf{Analytical Synthesis}: LLM invokes checked calculation tools for debt-service
    coverage ratios and stress scenarios against the forecasts.
    Human assesses the realism of management's assumptions.
  \item \textbf{Evaluative Judgment}: Human issues the opinion,
    structured by the four-stage ACP workflow. LLM acts as an
    adversarial reviewer raising counter-arguments.
  \item \textbf{Professional Communication}: Human authors the
    going-concern memorandum. LLM assists with footnote
    cross-references and PCAOB-citation formatting.
\end{itemize}
The protocol is designed for an evaluative-judgment task under
ambiguity; it explicitly requires adversarial review to defend
against the anchoring risk documented in the automation-bias
literature \citep{skitka2000,goddard2012}.

\subsubsection{Expected Credit Loss (ECL) Estimation (Reporting Domain)}

The illustrative ECL case uses an IFRS~9 loan-portfolio setting under the
general approach. Probability of default,
loss given default, and exposure at default support the calculation; assessment
of a significant increase in credit risk determines the applicable staging.
US GAAP credit-loss requirements must be instantiated separately rather than
treating IFRS~9 staging as an ASC~326 requirement.
The ACP specifies:
\begin{itemize}[leftmargin=2em]
  \item \textbf{Data Acquisition}: LLM extracts loan-level
    attributes and historical default data. Human reviews
    for completeness.
  \item \textbf{Analytical Synthesis}: LLM assists execution and interpretation of a specified
    quantitative model using checked tools. Human
    assesses model adequacy and identifies qualitative
    adjustments (the ``overlays'' common in practice).
  \item \textbf{Evaluative Judgment}: Human authors the
    significant-increase-in-credit-risk (SICR) determination
    for each loan and the overlay justification. LLM
    challenges the consistency of overlay applications across
    loans.
  \item \textbf{Professional Communication}: Human authors the
    IFRS~7 credit-risk disclosures. LLM assists with cross-checking
    internal consistency.
\end{itemize}

\subsubsection{Level-3 Fair-Value Measurement (Valuation Domain)}

For Level-3 (unobservable-input) fair-value estimates, the ACP
specifies:
\begin{itemize}[leftmargin=2em]
  \item \textbf{Data Acquisition}: LLM collects market
    comparables, prior transaction data, and relevant
    macro variables.
  \item \textbf{Analytical Synthesis}: LLM assists execution of checked
    valuation models (DCF, market multiples, prior-transaction)
    and presents a fan of estimates.
  \item \textbf{Evaluative Judgment}: Human selects the
    point estimate and defends the weighting of inputs.
    LLM presents an adversarial case for alternative weightings.
  \item \textbf{Professional Communication}: Human authors
    the Level-3 disclosure including the applicable unobservable-input
    sensitivity disclosures.
\end{itemize}

\subsection{Failure-Mode Coverage}

The ACP is designed to defend against five specific failure modes
documented in the literature.
Table~\ref{tab:failures} maps each failure mode to the protocol
element intended to address it. Coverage is a design mapping, not measured mitigation.

\begin{table}[H]
  \singlespacing
  \centering
  \caption{ACP Coverage of Documented Failure Modes in Human-AI
    Collaboration.}
  \label{tab:failures}
  \begin{tabular}{@{}>{\RaggedRight\arraybackslash}p{3.2cm}>{\RaggedRight\arraybackslash}p{4.4cm}>{\RaggedRight\arraybackslash}p{4.6cm}@{}}
    \toprule
    \textbf{Failure Mode} & \textbf{Description} & \textbf{Proposed ACP response} \\
    \midrule
    Anchoring
      & Premature convergence on AI suggestion
      & Stage A independent estimate before AI exposure \\
    \addlinespace
    Automation bias
      & Acceptance of AI output without verification
      & Stage C structured verification questions \\
    \addlinespace
    Miscalibrated trust
      & Over- or under-reliance due to
        uncalibrated confidence
      & Stage B explicit assumptions + source links \\
    \addlinespace
    Hidden rationale
      & Cannot defend AI conclusion under partner review
      & Stage B structured rationale panel \\
    \addlinespace
    Omitted evidence
      & AI suggestion anchored on incomplete evidence
      & Stage C Q2 evidence-addition prompt \\
    \bottomrule
  \end{tabular}
\end{table}

\section{Propositions and competing explanations}\label{sec:propositions}
The allocation matrix motivates conditional propositions; it does not logically
prove a performance advantage. The following claims separate the proposed
mechanisms from the design constraints. All are untested in accounting within
this paper. A sufficiently precise estimate in the opposite direction would
challenge the relevant proposition. A nonsignificant estimate alone would not
falsify it or demonstrate equivalence.

\paragraph{P1: verified information synthesis}
When the model has access to relevant evidence, its output is independently
checkable, and the human actually verifies material claims, assisted synthesis
is expected to reduce reference-range deviation relative to unaided synthesis.
Poor source coverage, calculation errors, or verification costs can reverse
this prediction. This is a capability-contingent prediction, not a claim that
all numerical tasks favor an LLM.

\paragraph{P2: harmful reliance under unresolved evaluation}
When advice is erroneous or poorly supported and the evidence leaves room for
interpretation, unstructured exposure is expected to increase movement toward
the advice that worsens reference agreement. Algorithm aversion and well-founded
rejection provide competing predictions. Movement toward sound advice is
appropriate updating, not harmful anchoring; the direction of movement alone
cannot adjudicate P2.

\paragraph{P3: the net effect of a required checkpoint}
The net effect of assigning a checkpoint requiring explicit comparison of
source evidence and reasoning is expected to improve final reference agreement relative to otherwise identical AI access.
The mechanism is additional scrutiny of material claims. Ceremonial completion,
reactance, and attachment to an initially wrong human estimate can erase or
reverse the benefit. P3 concerns the total effect of the required checkpoint,
including its attention and time costs, not an isolated psychological mechanism.

\paragraph{P4: the total effect of offering refinement}
The total effect of offering structured challenges and model responses after
verification is expected to improve reference agreement beyond verification alone. The
benefit may be larger where assumptions genuinely remain contestable, but
additional persuasive errors or time pressure may reverse it. Offering
refinement and actually using it are distinct interventions; the proposed
randomized contrast estimates the effect of offering it.

\paragraph{P5: a task-contingent cost}
Required verification is expected to increase professional time relative to
unstructured access, while the value of the accuracy--time trade-off varies
with task verifiability and ambiguity. Faster access to evidence could offset
the added steps, so a net time increase is not guaranteed. Accuracy and time
must first be reported separately; an incremental accuracy-per-minute ratio
is uninterpretable when the incremental time is near zero or negative.

Calibration is a separate \emph{measurement and implementation requirement},
not an efficacy proposition. A study holding uncertainty disclosure constant
cannot identify the causal effect of that disclosure. Similarly, the proposed
study holds independent estimation constant and therefore cannot estimate the
benefit of Stage~A itself. Testing those components requires additional
randomized conditions.

The candidate human evaluation below would reserve confirmatory inference
for P3 and P4 only after its measurement and feasibility gates are met.
P1, P2, and P5 remain exploratory because a three-case design cannot cleanly
separate domain, demand, advice quality, and case idiosyncrasies. A convincing
test of CDAJ's allocation predictions ultimately needs multiple independently
sampled tasks per cell and factorial variation in demand and advice quality.

\section{Human-evaluation agenda}\label{sec:innovation3}
The conceptual propositions require future human evidence. The public-source
experiment and synthetic validators cannot establish better professional
judgment, and the planning calculations below do not validate a behavioral
endpoint. The technical supplement preserves a candidate four-condition design:
unaided human judgment (C1), fixed model advice (C2), required verification (C3),
and verification plus optional refinement (C4). All initial estimates would be
collected before assignment and exposure. A possible confirmatory family is
K2=C3$-$C2 and K3=C4$-$C3, interpreted as total interface effects including
attention, time, and dialogue rather than isolated cognitive mechanisms.
Other contrasts and case-level effects remain exploratory.

Several gates must precede any confirmatory registration. A proposed independent
five-person panel would rate at least 12 evidence/assumption/justification items
per case and document defensible range boundaries and disagreement. Ordinal
agreement analysis \citep{krippendorff2011} would concern those item ratings,
not a single judgment or a fabricated reliability coefficient. An independent
pilot must then examine distributional floor effects, scale stability, advice
quality, and whether the endpoint reflects the intended judgment. One candidate
endpoint, normalized distance outside a defended range, measures admissibility
agreement and deliberately does not rank answers inside the range. It is not
assumed to be Gaussian or already validated. The supplement specifies separate
reasoning-quality and justified central-estimate sensitivity measures.

Professional minimum important improvements, acceptable added time, joint-power
requirements, recruitment access, burden, compensation, and budget must be
established before selecting a sample size. The numerical scenario of 225
complete-case-equivalent participants per arm is a hypothetical illustration,
not a funded or feasible recruitment commitment. Local ethics determination,
consent, privacy and third-party processing arrangements must precede human
recruitment. No reference panel, effectiveness pilot, ethics approval, preregistration, or
professional effectiveness-study sample is claimed for this proposed experiment. If these measurement and feasibility
gates cannot be met, the next study should be a feasibility investigation rather
than a nominally confirmatory test. Full candidate procedures, scoring options,
missingness assumptions, and registration prerequisites are retained in the
supplement so that this limitation remains inspectable.

\section{Public-source ambiguity diagnostic and computational verification}\label{sec:evaluation}
\subsection{Scope and reproducibility}
We executed a public-disclosure source-ambiguity diagnostic, event-order and
record-integrity checks, and prospective Gaussian planning experiments. \texttt{gpt-6-astra-max} (model display name used for the
OpenAI model accessed through the configured service)
assisted preparation and checking of the Python code;
the deterministic scripts, rather than model assertions, generated the reported
retrieval, check, and simulation metrics. The computational results in this section are distinct from the proposed
human effectiveness study and the separate coding exercise reported below. The accompanying reproducibility archive records source URLs and hashes,
limited source-derived cards, scripts, environment, seeds, derived retrieval
scores, simulation draws, machine-readable results, and checksums.
Publicly reported financial disclosures are real source data; the synthetic
fixtures and Gaussian draws are generated test inputs. Neither is a participant
response dataset. No accounting recommendation from an experimental LLM run or
practitioner effectiveness evaluation was used in these computational checks.
The programs check supplied mode/event and record contracts; they do not
implement an automatic CDAJ task allocator. The checks establish selected implementation
and statistical properties under stated assumptions; they do not validate
CDAJ's allocation rules or ACP's behavioral effectiveness.

\subsection{A public-disclosure provenance and ambiguity diagnostic}\label{sec:public}
\subsubsection{Question and corpus construction}
The diagnostic question is whether a particular page-level co-occurrence rule
changes recovery of a designated source page, and what ambiguity remains
between page recovery and the source's printed table labels. It is not a test
of structured context requirements in general.
This exercises one evidence-location operation relevant to ACP, not the
protocol's effect on people or the professional correctness of an accounting
interpretation. The primary comparison supplies \emph{identical
query information} to ordinary bag-of-fields retrieval and a deterministic
context-presence gate. A value-only search is a secondary information-removal
ablation, not a realistic analyst or LLM baseline.

We downloaded the complete issuer-hosted 2024 annual reports of HSBC Holdings,
Barclays, and Standard Chartered on 22 November 2026
\citep{hsbc2025report,barclays2025report,sc2025report}. The named issuer/year
selection, eligible cells, scoring methods, and amendments were recorded before
retrieval scoring. All three downloads succeeded without substitution.
The reports contain 460, 536, and 410 PDF pages respectively, totaling 1,406;
three text-empty pages remain in the corpus. Source URLs, download timestamps,
PDF hashes, extraction-version information, and failures/exclusions are retained.
This is a purposive, evaluation-only feasibility corpus, not a random sample,
training set, or held-out test of a fitted model.

Two consolidated topic tables per report were selected: credit impairment/ECL
and Level-3 financial assets. Within each designated table, the first three
eligible 2024 numeric cells in row-major order were retained, restricted to the
stated topic columns. Dashes were not numeric cells; two were skipped and
logged. Repeated amounts remained separate row/column observations. The resulting
18 cards cluster within six source pages and three issuers. Each card preserves
the report, PDF and printed page, table, row, column, period, unit, and printed
value. \texttt{gpt-6-astra-max} assisted extraction and visual checking of the six pages;
a second agent checked the same source cells. The initial labels were model-assisted, not independently generated human
annotations or practitioner judgments. Subsequently, the human requester confirmed completion of checks of all
18 source values, row/column labels, periods, and units using a source packet;
the same confirmation covered the five post-hoc comparisons reported below. No corrections were requested, and the frozen
labels and results were unchanged. The actual checker's identity, professional qualifications, and blinding were
not separately supplied.
This is a documented human source check, not independent expert adjudication.
A later technical cross-check used a different PDF extraction library,
\texttt{pdfplumber} 0.11.9, against the same hashed reports: all 18 printed
values and exact unit strings were found on their designated pages, with
multiple value occurrences for eight cards. Coordinate witnesses and ambiguity
are retained. This independently implemented extraction corroborates text
presence; it does not assign a value to a particular row/column or constitute
independent human adjudication.

\begin{table}[H]\centering\singlespacing\small
\caption{Public-source corpus and designated reference pages.}
\label{tab:corpus}
\begin{tabularx}{\textwidth}{@{}Xrrrr@{}}
\toprule
Issuer / 2024 annual report & Total PDF pages & ECL origin & Level-3 origin & Cards\\\midrule
HSBC Holdings & 460 & 144 (142) & 389 (387) & 6\\
Barclays & 536 & 308 (307) & 479 (478) & 6\\
Standard Chartered & 410 & 212 (210) & 325 (323) & 6\\\bottomrule
\end{tabularx}
\par\smallskip\footnotesize Note: Origins are one-based PDF pages, with printed
page numbers in parentheses. Three cards per topic table; these are clustered
cells, not 18 independent reports or source pages.
\end{table}

\subsubsection{Frozen retrieval methods and metrics}
All methods search the complete combined page corpus. Queries contain only the
printed value, metric phrase derived from row/column labels, period, and unit.
Exact origin IDs, page locators, and issuer identity are withheld from the query.
Because these fields are constructed from the source itself, the task is a
favorable controlled-query test, not a sample of naturally occurring requests
or a test of generating the correct context.

Method B is ordinary BM25 with fixed $k_1=1.2$ and $b=0.75$
\citep{robertson2009}. It uses each distinct query token once and the positive
IDF form $\log(1+(N-df+0.5)/(df+0.5))$. Method C uses exactly B's scores and query
but retains only pages containing the canonical value and every nonempty
period and unit token; it abstains if none are eligible. Method A, the secondary
ablation, ranks pages by frequency of the value token alone. There is no learned
model, manual reranking, result-dependent tuning, or source-page hint.

The fixed tokenizer lowercases NFKC-normalized text, canonicalizes comma-grouped
numbers and decimals, and treats parenthesized numeric literals as negative.
It retains ASCII letter and numeric tokens with different prefixes but removes
currency symbols: both \texttt{\$m} and \texttt{\pounds m} yield the word token
\texttt{m}. This explicitly limits currency discrimination. It does not reconstruct
table coordinates or verify that a year/unit governs the matched value rather
than a different cell on the same page. Exact preprocessing, code/input hashes,
pre-scoring amendments, and numerical checks are supplied.

The primary descriptive metrics are designated-origin-page Recall@1 and
Recall@3. Reciprocal rank and candidate-page counts are secondary. A fixed
report/page order breaks score ties for displayed ranks; tie-averaged metrics
also average over all placements within a tied group. An origin with zero score
or excluded by C receives zero recovery, even if deterministic zero-score
ordering would place it near the top. All cards remain in the results.
No population confidence interval or significance test is reported: the 18
purposively selected, clustered cells are not independent sampled firms.

\subsubsection{Results: no incremental Recall@1 or Recall@3 benefit from the gate}
\begin{table}[H]\centering\singlespacing\small
\caption{Designated-source-page recovery on the 18 public disclosure cards.}
\label{tab:retrieval}
\begin{tabularx}{\textwidth}{@{}Xrrrr@{}}
\toprule
Method & Recall@1 & Recall@3 & Mean recip. rank & Tie-avg. Recall@1\\\midrule
A: value-only ablation & 7/18 & 11/18 & 0.5200 & 0.2545\\
B: full-field BM25 & 13/18 & 18/18 & 0.8333 & 0.7222\\
C: same-information gate & 13/18 & 18/18 & 0.8426 & 0.7222\\\bottomrule
\end{tabularx}
\par\smallskip\footnotesize Note: B versus C is the primary comparison; A
removes information. B and C have identical deterministic and tie-averaged
Recall@1/@3 here. Exact-source recovery is not semantic accounting accuracy.
\end{table}
Method C can only delete candidates; it does not change the relative order
of pages it retains. A recovery gain requires removing a higher-ranked
distractor. The comparison therefore concerns this limited co-occurrence rule,
not the value of all structured evidence verification.

The added gate produced no Recall@1 or Recall@3 gain on any card: all 18
comparisons were unchanged. One origin moved from rank three to rank two,
explaining the small reciprocal-rank change; this is not evidence of general
retrieval superiority. No query caused gate abstention or exclusion of its
origin. Thus the experiment found no incremental Recall@1 or Recall@3 benefit from merely
requiring value/period/unit co-occurrence beyond full-field BM25.

B's Recall@1 was 5/6 for HSBC, 5/6 for Barclays, and 3/6 for Standard Chartered;
by topic it was 8/9 for ECL and 5/9 for Level~3. Compared with A, B improved
tie-averaged Recall@1 on ten cards, left five unchanged, and worsened three;
all three losses were Standard Chartered Level-3 cards. The mean number of
pages containing the value was 35.83 overall, 68.11 for ECL and 3.56 for
Level~3. These counts describe this selection of small repeated and larger
amounts; they do not establish a population difference between accounting tasks.
All 18 source values were found on their recorded pages. The archive retains
all 25,308 card-by-page score rows and each card's ranks, ties, eligible pages,
and source-token checks.

\subsubsection{Post-hoc label comparison for the five origin misses}
After scoring was frozen, an explanatory visual audit examined the five
non-origin top-ranked candidates, on four unique pages. Three visibly repeated
the reported closing balances: HSBC derivative assets of 2,149; Standard
Chartered customer loans of 1,937; and Standard Chartered debt securities of
1,593, with a slightly expanded category label in the alternative table.
A fourth repeated Standard Chartered's reverse-repurchase amount of 3,239 in
a sensitivity table labeled net exposure, rather than the origin's asset
fair-value hierarchy. That presentation difference prevents automatic semantic
equivalence.

The fifth candidate has different literal table and column labels from the
requested source cell. The target
was Barclays' 2024 UK retail-mortgage Stage~1 impairment allowance of
\pounds36 million (PDF page~308). The retrieved page~324 contained a retail-mortgage
economic-uncertainty-adjustment table: its \emph{total} was \pounds36 million,
whereas its Stage~1 amount was \pounds7 million. Value, year, unit, and row words
could therefore co-occur while the value belonged to a different measure and
column. The frozen gate did not resolve this mismatch.

These comparisons concern the reports' printed amounts and labels. They do
not adjudicate the correctness of the issuers' accounting treatment, the
appropriateness of an estimate, or how a professional should use the evidence.
The checker's qualifications were not reported, and no expertise-based
conclusion is inferred from the confirmation.

This post-hoc audit neither relabels the provenance targets nor produces a new
semantic accuracy rate. Only the five origin misses were inspected for this purpose;
there is no prespecified, independently adjudicated semantic reference set.
The original 13/18 and 18/18 metrics remain unchanged. The audit instead exposes
two distinct limits: a non-origin page need not be wrong, and a token-matching
page need not support the intended accounting claim. No going-concern empirical
case was constructed, and no participant or LLM accounting-performance outcome
was measured.

\paragraph{An elementary information-loss observation}
Let $R(s)$ retain only a source page's pooled token counts and length, and let
$q$ be the claim that portfolio A's displayed allowance is 36. Consider two tables with identical portfolio labels A and B and an
allowance header, but with the values 36 and 7 exchanged between the rows.
Keep all headings, period/unit labels, and the rest of the corpus fixed.
Then $R(s_1)=R(s_2)$, but $q$ is true for one page and false for the other. A verifier observing only $(R(s),q)$ must produce
the same decision, or randomized decision distribution, in both cases.
It therefore cannot guarantee both no false acceptance and acceptance of
every true claim for arbitrary such printed-cell claims, with probability one
in the randomized case. Abstention can protect soundness by foregoing
completeness; useful approximate performance is not ruled out. This elementary observation concerns
loss of row-to-value binding, not correctness of accounting treatment, a bound
on actual retrieval rates, or evidence that humans or ACP solve the problem.
The observed corpus results remain the descriptive results above.

The design implication is a requirement for \emph{claim-level table binding}:
for tabular evidence, a locator should preserve document/version, page, table,
row, column, period, unit, and literal value. The public cards supply these
fields. The base record checker verifies identifier and locator presence. A later
typed-contract extension also checks consistency with a supplied cell-binding
catalog and value/period occurrence, but it does not parse a table or establish
semantic entailment. The source-label diagnostic motivates that additional verification obligation
and delimits a weak implementation, rather than validating ACP's human checkpoint.

\subsection{Selected event-order and payload checks}
A new reference validator checks a limited set of one-case ACP event traces.
Five valid synthetic traces and 21 deliberately mutated traces test such
conditions as missing stages, condition-inappropriate events, a mismatched
frozen-output hash, an unmatched refinement response, and finalization during
a pending exchange. The validator returned the intended verdict on all 26
fixtures. This is an author-designed software correctness exercise, not
independent validation against real sessions.

The validator does not check the presence or substantive quality of actual
initial estimates, initial evidence, final conclusions, or signatures; nor
does it verify the quality of explanations or verification responses. It
does not implement a browser interface, identity controls, append-only storage,
privacy, or isolation across participants and cases. Its strict timestamp order
is a simplifying assumption; a production system should use a reliable event
sequence because real millisecond timestamps may tie. Passing these tests shows
that selected workflow constraints can be encoded, not that an ACP system is
ready for deployment or that users meaningfully verify advice.

\subsection{Finalized-record integrity checks}
A separate executable validator implements the record rules in
Table~\ref{tab:record}. It obtained the expected decisions on 33 synthetic
fixtures: two acceptable records (with and without an optional exchange) and
31 deliberately invalid mutations. The mutations include a missing initial
value or final conclusion, changed source content with an unchanged hash,
dangling evidence or assumption IDs, an untracked contested claim, missing or
duplicate exception dispositions, and missing or unconfirmed sign-off. Complete
inputs, decisions, and rejection reasons are supplied in the archive, together
with the explicitly fabricated accounting specimen described above.
A rerun reproduced the four JSON result/fixture outputs byte for byte.

The new checker verifies presence and cross-references, not truth or professional
adequacy. A source hash cannot authenticate a document if an attacker controls
both the document and its hash. A declared ``supported'' label, an exception
marked ``addressed,'' and a synthetic sign-off flag can all be false while the
record remains schema-conforming. The checker does not verify event order;
the event checker does not verify record content. Their test counts must not be
combined into a certification of an integrated ACP platform. The substantive
result is narrower: the stated evidence-to-exception-to-conclusion relations
are executable requirements rather than only narrative recommendations.

\subsection{Typed-contract extension and a deliberate semantic counterexample}
A separately versioned extension retains the original 33-fixture validator
unchanged and adds the typed accounting-context and source-binding rules.
Its 42 deterministic fixtures produce the specified structural decisions:
40 rejections and two acceptances. Mutations test missing fields, invalid ranges,
dangling catalog/source references, wrong measures or columns against an
unchanged requested binding, inconsistent literal normalization, and missing
value/period occurrences despite matching source hashes. Five JSON outputs
reproduce byte for byte in the recorded runtime. These are software fixtures,
not sampled accounting errors, and no error-detection rate for professional
work is inferred.

One accepted fixture is a fully labeled fabricated record with unresolved
guidance and an unestablished admissible range; its professional conclusion is
withheld. The other is a \emph{deliberate semantic counterexample}: the claim's
free text still requests a Stage~1 impairment amount while both the structured
requested binding and supplied binding are changed to an adjustment total.
The self-consistent typed contract passes although the free-text meaning
conflicts. By contrast, changing only the supplied binding while retaining the
original requested identity is rejected. The fixture is inspired by the real
same-36 disclosure collision but its catalog and record text are fabricated,
not additional report observations or expert labels.

This result shows both the value and the boundary of the formal artifact.
A typed contract can reject inconsistent source substitution relative to a
fixed request; it cannot decide which request, catalog entry, or accounting
interpretation is substantively correct. The accepted counterexample is
reported as a limitation of semantic assurance, not counted as a correct
accounting outcome. No original public-source label, retrieval score, or
version-one fixture was changed for this extension.

\subsection{Planning illustration under a hypothetical continuous endpoint}
This illustration concerns a hypothetical continuous endpoint; it is not
a power validation for the unpiloted range-distance score. Assume three
equally weighted observations per participant,
equal single-case variance $\sigma^2$, and compound-symmetry correlation
$\rho$. The variance of a participant mean is
\[
 \operatorname{Var}(\bar Y_i)=\sigma^2\frac{1+2\rho}{3}.
\]
For equal arm size $n$ and a standardized difference in units of the single-case
standard deviation, the standard error is
\[
 SE_d=\sqrt{\frac{2(1+2\rho)}{3n}}.
\]
At $n=225$ and $\rho=0.30$, $SE_d=0.068853$. At the conservative first
Holm threshold \citep{holm1979}, two-sided $\alpha=0.025$, the
normal-approximation 80\% minimum detectable difference is
\[
 d_{80}=(z_{0.9875}+z_{0.80})SE_d=0.212276.
\]
Fixed-threshold power for $d=0.20,0.30,0.40$ is 0.7464, 0.9828, and 0.9998,
respectively. These are assumption-dependent planning quantities, not achieved
precision. The corresponding one-case MDE is 0.2907, illustrating why pooled
power cannot be transferred to case-specific claims. Raising $\rho$ from 0 to
0.60 raises the pooled MDE from 0.1678 to 0.2489. A more flexible covariance
model is not guaranteed to be more powerful.

The fixed $n=225$ target is retained as a planning scenario, not justified by a
universally meaningful 0.40 effect or a demonstrated recruitment capacity.
A practically important improvement should be agreed with practitioners on the
raw rubric scales before a future study fixes its sample size. The formulas
assume common variance; pilot-based normalization does not guarantee that
assumption across arms or in a new sample.

\subsection{Seeded Monte Carlo results}
Seven scenarios used 100,000 repetitions each, with seeds 20261122--20261128.
Each repetition simulated Gaussian arm means for C2, C3, and C4 with variance
$v(1+2\rho)/(3n)$, where $v=1$ except in the doubled-variance stress scenario, calculated the two contrast statistics using the known
variance, and applied Holm adjustment. C1 is unnecessary for these two tests.
Because C3 is shared with opposite signs, contrast errors have correlation
$-0.5$ under this model; this is an imposed algebraic property, not a behavioral
finding.

\begin{table}[H]
\centering\singlespacing\small
\caption{Operating properties under the declared Gaussian planning model.}
\label{tab:simulation}
\begin{tabularx}{\textwidth}{@{}Xll@{}}
\toprule
Scenario & Quantity & Estimate [95\% MC interval]\\\midrule
Both increments zero & Any false rejection & 0.04662 [0.04533, 0.04794]\\
One null; other $-0.40$ & Rejection of true null & 0.05039 [0.04905, 0.05176]\\
Both increments $-0.20$ & Both rejected & 0.65671 [0.65376, 0.65965]\\
Both increments $-0.30$ & Both rejected & 0.98296 [0.98214, 0.98374]\\
Both $-0.30$; $\rho=0.60$ & Both rejected & 0.92212 [0.92044, 0.92376]\\
Both raw increments $-0.30$; variance doubled & Both rejected & 0.73691 [0.73417, 0.73963]\\
\bottomrule
\end{tabularx}
\vspace{4pt}
\parbox{\textwidth}{\footnotesize Note: Wilson intervals \citep{wilson1927} quantify Monte Carlo
sampling uncertainty, not uncertainty about real accountants. Default
$n=225$, $m=3$, $\rho=0.30$, single-case variance 1. The last row holds raw
effects fixed while increasing variance, so standardized effects decrease.
The seventh scenario, both increments $-0.40$, and all raw draws are in the archive.}
\end{table}
The estimated global-null familywise error is below the nominal 0.05 level.
With both true increments $-0.20$, Holm-adjusted marginal rejection probabilities are 0.82215
and 0.81585 (distinct from the fixed-0.025 power of 0.7464), but joint rejection is only 0.65671. The distinction matters when
a theory requires both verification and refinement to add value. The partial-null
result of 0.05039 is compatible with 0.05 within Monte Carlo uncertainty;
it is not evidence of error control above its nominal bound.

A full rerun reproduced 16 archived output files byte for byte.
A separate verification script reread the simulation draws and checked sampling
variances, contrast correlation, and fixed-threshold rejection probabilities
against the analytic model. These checks validate the implementation of the
specified planning experiment. They do not model estimated variances,
non-normal range distances, floor effects, dropout, noncompliance, or the exact
finite-sample behavior of the proposed mixed model. Pilot-informed simulation
of those features remains necessary before recruitment.

\section{Primary coding exercise and execution deviation}\label{sec:human-coding}
\subsection{Materials, returns, and scope}
Two separately completed response workbooks were received for 20 frozen,
constructed hypothetical accounting case descriptions. The cases deliberately
include compound, boundary, and underspecified descriptions; they are neither a
random task sample nor observations of professional engagements. They differ
from the 25 illustrative allocations in Appendix~\ref{app:matrix}. Procedural instructions were in Chinese; definitions and case descriptions
were bilingual Chinese--English, and financial-report sources were in English. No preferred labels were supplied for the 20 cases. Coders could mark
a description as SINGLE, request SPLIT, or mark its boundary UNCERTAIN. Stage
codes refer to CDAJ's four process stages, not ACP's four interface stages.
Demand M retains both information-intensive and evaluative demands; U is an
administrative uncoded marker, not a fifth stage or a new demand type.

Both coders declared consent to research use, distinct personhood, independent
completion, no prior design involvement or exposure to the paper or proposed
test-item labels, no consultation with the other coder, and no generative-AI
assistance with judgments. One reported working as an accountant with a
bachelor's degree in accounting; the other reported being an audit manager and
accounting PhD student, with a Chinese CPA qualification and five years of
Big Four audit experience. These are self-reported roles and qualifications,
not authenticated credentials. Both reported no relationship to the authors
and no support or compensation. Reported language use was English and Chinese--English. Comprehension of the
Chinese instructions was not separately assessed. Both completion and
primary-lock dates were 22 November
2026. The authors report that Jiangxi University of Finance and Economics confirmed on
22 November 2026 that ethics review was not required; no reference number was
provided. The specific office, response form, and basis have not been supplied.
No institutional document has been independently verified, and no prior approval
is inferred from this report. Individual responses and identifying metadata remain private; only
aggregate statistics are reported.

Original workbooks were hashed and preserved. Audited working copies normalize
one case-only human-origin declaration and remove two ID-only, otherwise blank
duplicate placeholder rows where a completed row exists. These administrative
repairs do not change a unitization, stage, or demand response. All initial
primary labels remain unchanged; no consensus or adjudicated labels enter the
analysis.

\subsection{Unresolved auxiliary formats and limited execution}
The frozen full-plan reader stopped with a format error. Both returns contain
numeric realism responses, whereas the distributed guide and workbook specified
PLAUSIBLE, IMPLAUSIBLE, or UNCLEAR. The original workbook had category dropdowns
but disabled input and error alerts, a weakness in the response interface. The
returned validation settings differ from the original; the available files do
not establish why. No numerical-to-category conversion is justified. A source
comparison response also contains two categories in a field specified for one.
These auxiliary answers are preserved without recoding or a preferred
interpretation. The complete frozen analysis therefore remains unfinished.

Realism and source-comparison answers do not enter the declared primary coding
estimands or coder-declaration eligibility rules. For this primary-only manuscript revision, a
separate execution scope was documented on 22 November 2026 at 08:33:52 UTC,
before the first agreement-coefficient call at 08:34:40 UTC. It validates the primary fields for the same two coders and all
20 cases and uses the frozen agreement calculation and comparability rules.
This is a disclosed execution deviation after returns were available, not a
public preregistration or a claim that the complete original pipeline passed.
No coder or case is removed because of the auxiliary formats, and no missing
or unresolved answer is imputed.

\subsection{Reporting rules}
The mandatory descriptive layer retains all 20 cases, including split requests,
uncertain boundaries, U, M, and missing responses. Exact raw agreement includes
explicit unresolved labels and consequently does not by itself show useful
classifiability. Secondary nominal Cohen's $\kappa$ is calculated separately
for each axis only where both coders select SINGLE and both provide a nonmissing,
non-U code. The demand coefficient compares I/E/M response profiles; it is not
an exclusive binary I/E coefficient or validation of an eight-cell taxonomy.
Coverage of resolved I/E cases is therefore reported separately. Unaligned
subunits are not pooled for agreement. There are no population confidence
intervals, significance tests, passing thresholds, or consensus coefficients.

\subsection{Aggregate primary results}
All 20 cases have paired unitization, stage, and demand responses; none is
missing. The two anonymous response profiles contain respectively 14 and 15
SINGLE decisions, three and two SPLIT decisions, and three UNCERTAIN decisions
each. Exact agreement on the full unitization--stage--demand triple is 17/20.
Table~\ref{tab:human-coding} reports the all-case and conditional results.

\begin{table}[H]
\centering\singlespacing\small
\caption{Primary coding agreement and coverage for 20 constructed cases.}
\label{tab:human-coding}
\begin{tabularx}{\textwidth}{@{}Xrrr@{}}
\toprule
Quantity & Unitization & Stage & Demand\\\midrule
Complete paired responses & 20/20 & 20/20 & 20/20\\
All-case exact agreement & 19/20 & 20/20 & 18/20\\
All-case agreement rate & 0.950 & 1.000 & 0.900\\
Conditional comparable cases & --- & 14/20 & 14/20\\
Conditional exact agreement & --- & 14/14 & 12/14\\
Conditional nominal $\kappa$ & --- & 1.000 & 0.752\\
\bottomrule
\end{tabularx}
\vspace{4pt}
\parbox{\textwidth}{\footnotesize Note: All-case agreement includes U and
labels supplied on non-SINGLE parent descriptions. Conditional results retain
only jointly SINGLE cases with nonmissing, non-U axis codes. The demand
coefficient concerns I/E/M profiles, not exclusive I/E classification. No
unitization coefficient was specified.}
\end{table}

For both axes, six cases are excluded from the conditional coefficient because
they are not jointly SINGLE; three of those also carry explicit U codes.
These are overlapping exclusion reasons, not nine exclusions. Within the 14
comparable cases, chance-expected agreement from the observed marginals is
0.265306 for stage and 0.423469 for demand. Across all 20 descriptions, each
profile uses stage labels S1 five times, S2 four times, S3 four times, S4 four
times, and U three times. Demand frequencies are I/E/M/U = 5/7/5/3 and
5/8/4/3 for the two anonymous profiles. Some raw stage or M labels occur on
non-SINGLE parents; they are retained descriptively but excluded from the
conditional calculations. The frozen structural checks also flag non-SINGLE
parent axis labels in three descriptions in one profile and two in the other.
For two SPLIT descriptions, one profile supplies fewer than two subunit rows.
These adverse diagnostics are retained without repairing labels, aligning
subunits, or introducing further exclusions; the six non-jointly-SINGLE cases
are already excluded by the original conditional rule.

Individually, 12/20 and 13/20 descriptions are SINGLE with an I or E demand;
each profile also has two SINGLE descriptions marked M. Jointly resolved
I/E coverage is 12/20 (60\%). One case is jointly SINGLE and M for both coders;
two jointly SINGLE cases have M from at least one coder. Accordingly, perfect
conditional stage agreement coexists with unresolved unit boundaries and
incomplete exclusive-demand coverage. The findings show repeatability of some
coding responses under these materials and declarations. They do not establish
correct labels, natural-task coverage, optimal human--AI allocation, or ACP's
effect on judgment. No optional allocation responses were supplied. The
source-check responses are analyzed separately below; they do not constitute
source accuracy scores or expert adjudication.

\subsection{Separate descriptive source-response exercise}\label{sec:source-responses}
Both returns also contain responses to all 18 source-check items, yielding 36
records. A separate execution scope was documented on 22 November 2026 at
09:27:45 UTC, before the first source-status aggregation at 09:28:42 UTC. It
retains both respondents and all 18 items, and describes the five component
fields and overall judgment under their original schemas. It neither changes
the primary coding rules nor treats the original full-plan error as resolved.
The 40 numeric realism answers and the unresolved fifth comparison remain
outside this descriptive execution; no new coefficient or passing threshold
was introduced.

\begin{table}[H]
\centering\singlespacing\small
\caption{Pooled recorded source-check statuses across 36 response records.}
\label{tab:source-responses}
\begin{tabularx}{\textwidth}{@{}Xrrrr@{}}
\toprule
Field & MATCH & DISCREPANCY & UNCLEAR & Missing\\\midrule
Value & 28 & 4 & 4 & 0\\
Row & 30 & 4 & 2 & 0\\
Column & 32 & 4 & 0 & 0\\
Period & 32 & 4 & 0 & 0\\
Unit & 32 & 4 & 0 & 0\\
Overall & 20 & 12 & 4 & 0\\
\bottomrule
\end{tabularx}
\vspace{4pt}
\parbox{\textwidth}{\footnotesize Note: Denominator is 18 items multiplied by
two returns. These are recorded judgments, not correctness scores against an
independently adjudicated reference. No item-level or respondent-linked
answers are disclosed.}
\end{table}

Sixteen of the 36 records have at least one adverse or unclear status. This
cannot be read as 16 distinct erroneous source cards: the counts refer to
responses, not independently confirmed defects. Component and overall statuses
are retained separately; an overall response is not used to fill or override
any component. The private audit retains every potential cross-field
inconsistency. A disclosure rule documented before aggregation suppresses
boundary counts that could reveal protected complete response labels when
combined with other tables. This privacy limitation is disclosed rather than
replaced with a claim that the source instrument passed.

The first four comparison items have complete schema-valid relationships.
Across their eight responses, each of four categories occurs twice: same
printed value and labels; different values; same value with different labels;
and related labels with unclear scope. Aggregate counts do not identify which
comparison or person received each label. The fifth item remains unresolved
because one answer combined categories; its omission from these counts is
explicit and does not turn a five-item instrument into a completed four-item
study. No relationship category is adjudicated as correct from these counts.

These responses supplement the initial requester confirmation with separately
returned source judgments, including adverse and uncertain judgments. They do
not resolve all source-label questions or authenticate professional expertise.
A source-location audit must bind the year, column, subgroup, and parent header
to the displayed value: repeated values or repeated leaf labels can occur
within a page. Such document checks are distinct from judging a respondent's
answer and must not overwrite the original response categories. The original
retrieval targets and all retrieval results remain unchanged, including the
absence of a Recall@1/Recall@3 gain and the single rank-three-to-rank-two move.

\section{Implications for AIS design and research}\label{sec:implications}
\subsection{System design and engagement review}
The public-source diagnostic gives a concrete limit to this weak implementation:
mandatory token co-occurrence added no recovery benefit at ranks one and three, and a retrieved
amount could appear under a different table or column label than requested. This supports treating
page retrieval as a candidate-evidence step, followed by explicit table-cell
and claim verification; it does not show that the proposed human checkpoint
actually performs that verification well.

The immediate design contribution is a specification of what an AI-assisted
judgment record should make inspectable: who supplied an estimate, which source
supports each material claim, what the human challenged, and why the final
conclusion followed. A system can require those fields and enforce event order
without demonstrating that they improve judgment. Managers should therefore
evaluate semantic quality and professional time alongside completion rates.
A fully populated checklist can conceal poor review.

CDAJ also suggests separating calculation services, retrieval, generative
rationale, and professional sign-off. Different components require different
checks. A correct deterministic calculation does not validate an unsupported
assumption; a credible citation does not prove that a source supports the
conclusion. Exception handling should allow a professional to suspend the AI
workflow when evidence is missing rather than force a nominally complete record.

If the proposed K2 contrast shows a precise benefit, it would support the
required checkpoint for the studied cases and outputs. If K3 adds no precisely
estimated practical benefit, refinement may warrant redesign or more targeted
use; a wide interval would instead leave its value unresolved. Neither outcome
would establish a general rule for all accounting work. Any deployment decision
must consider time, error severity, evidence access, and the quality of existing
review procedures.

\subsection{Professional oversight and education}
The protocol offers a candidate documentation structure for discussion with
firms, educators, and standard-setters, not a determination of regulatory
compliance. Evidence retention and accountable sign-off are relevant to review,
but the appropriate legal and professional requirements depend on the
jurisdiction and engagement. The computational checks cannot justify a mandate
to use ACP or a change in professional liability.

For education, the cases can structure exercises comparing initial judgments,
AI advice, source evidence, and revisions. The important teaching object is the
reason for accepting or rejecting advice. Improved learning or reduced anchoring
would need separate assessment; requiring students to explain disagreement is
not proof that they have acquired professional skepticism.

\section{Limitations and future research}\label{sec:limitations}
The principal limitation is the absence of human effectiveness evidence.
The public-disclosure diagnostic is also small and purposive: 18 source cells
in six tables across three banks, with query labels derived from the source.
It is not a representative accounting benchmark or a comparison of LLMs.
Source extraction and the post-hoc audit were AI-assisted and subsequently
confirmed in one human requester reply about the source packet. This initial check is not blinded expert adjudication. The later 36
source-response records include 16 with at least one adverse or unclear
status and do not establish an independently adjudicated reference standard.
Neither check justifies a revised semantic-success rate. CDAJ
is an author-constructed taxonomy, the allocation inventory is illustrative,
and the case files and scoring panel have not been finalized. These constraints
limit the contribution to conceptual specification, bounded technical
evaluation, and the reported coding repeatability exercise. The latter uses
constructed cases and conditional coefficients, with unresolved auxiliary
formats and no independently authenticated professional qualifications. It does
not validate allocation recommendations or professional usefulness. The future
effectiveness-study case files and scoring panel remain unfinalized.

The future design itself has limits. It estimates effects conditional on three
cases and frozen initial recommendations. It cannot disentangle cognitive
demand from domain or case difficulty, establish calibrated uncertainty,
isolate independent-estimation effects, or demonstrate complementarity relative
to AI alone. A multi-case factorial replication varying advice quality and
cognitive demand would be needed for the broader allocation propositions.
Cross-case carryover and professional heterogeneity require explicit treatment.

Reference-range distance is a tractable but incomplete proxy for judgment
quality. It can reward agreement with a mistaken panel, miss weak reasoning,
and produce many zero scores. Independent rubric validation and separate
reasoning-quality measures are essential. A richer computational evaluation
should use pilot-estimated outcome distributions and missingness, rather than
assuming that Gaussian power calculations transfer directly to bounded or
zero-inflated observations.

Finally, the protocol's burden can itself undermine its aims. A requirement to
state an independent estimate may fix an early misconception; verification may
become performative; and refinement may make an erroneous model more
persuasive. These are substantive alternatives to the proposed mechanisms.
Future evidence should be capable of rejecting or narrowing ACP's allocation
recommendations rather than treating all additional human involvement as a
benefit.

\section{Conclusion}\label{sec:conclusion}
CDAJ and ACP provide an accounting-specific way to connect task allocation,
source verification, professional evaluation, and a reviewable decision record.
The paper preserves human responsibility as a design constraint while treating
performance benefits as contestable propositions. A public-source diagnostic finds no benefit at ranks one and three from a simple context gate
and demonstrates why page-level matches do not verify a requested table-cell reference.
Reproducible record/event checks and planning experiments establish additional
limited technical properties and reveal the
precision requirements of a future evaluation. The next substantive test is
whether the specified increments improve defensible professional judgment enough
to justify their cost. That result remains to be established.

\section*{Data and code availability}
A separate reproducibility archive has been prepared and is maintained by the
authors. It contains official source URLs, acquisition/extraction code, PDF
hashes, limited source-derived cards, selection/failure logs, retrieval code and
scores, synthetic fixtures, simulation draws, and numerical checks. The computational archive and a separate supplement of anonymous primary-coding
and source-response aggregates are available in the same private repository
for reviewer access. The confidential review link is provided separately in
editorial materials. This is not a public deposit, and no registered DOI or
license grant is claimed.
The issuer reports and full extracted text are retained for audit but are not
redistributed here because an open redistribution license has not been
established. The cited official report URLs and version hashes support source
identification. This version also reports aggregate primary-coding and source-response
results from two returned forms. Original workbooks and derived individual responses are retained
privately; row-level sharing was not authorized by both coders. No human
effectiveness dataset has been generated.

\section*{Funding}
This research has not received a specific grant to date.

\section*{Declaration of competing interest}
The authors declare no competing interests.

\section*{Declaration of generative AI and AI-assisted technologies in the manuscript preparation process}
\texttt{gpt-6-astra-max} (model display name used for the
OpenAI model accessed through the configured service) was used
to assist manuscript revision, source-card extraction and visual checking,
methodological checking, and preparation of computational code and materials. DeepSeek-v4-pro (API response model identifier \texttt{deepseek-v4-pro},
DeepSeek) was used for independent automated critique of the revised manuscript and
package. These systems were not research participants or sources of observed
accounting behavior. No public weight hash or provider training-snapshot identifier was available
for the configured GPT deployment. The printed model name is a reader-facing display name; it is not a claim
of a separately verified public release or an exact API routing identifier. The authors are
responsible for verifying sources, code, interpretations,
and the final content. The initial AI-assisted source labels subsequently received a human confirmation of source checking; no independent expert annotation or professional adjudication is
claimed.

\appendix
\counterwithin{table}{section}
\renewcommand{\thesection}{\Alph{section}}
\titleformat{\section}{\normalfont\large\bfseries}{Appendix \thesection:}{0.5em}{}
\section{Allocation Matrix Detail}
\label{app:matrix}

Tables~\ref{tab:matrixdetail} and~\ref{tab:matrixdetail2} enumerate
the 25 illustrative accounting
work-unit examples and their proposed allocation hypotheses under the CDAJ
coding scheme.
Tasks are author-selected examples from audit, reporting, tax, and valuation.
They are proposed dominant codings of the named work units, not
practitioner-validated classifications of entire engagements. Broader tasks
must be decomposed using Section~\ref{sec:coding}; mixed units retain
alternative codings rather than forced labels.

\begin{table}[H]
  \singlespacing
  \small
  \setlength{\tabcolsep}{4pt}
  \renewcommand{\arraystretch}{1.05}
  \centering
  \caption{CDAJ Allocation Detail for 25 Illustrative Work Units
    (Part 1 of 2: Tasks 1--12).}
  \label{tab:matrixdetail}
  \begin{tabular}{@{}>{\RaggedRight\arraybackslash}p{4.0cm}>{\RaggedRight\arraybackslash}p{3.5cm}>{\RaggedRight\arraybackslash}p{7.5cm}@{}}
    \toprule
    \textbf{Work-unit example} & \textbf{Dominant Stage / Demand} & \textbf{Proposed Allocation} \\
    \midrule
    Weighing refinancing evidence (going concern)
      & Eval.\ / Evaluative
      & Human leads; LLM as adversarial reviewer \\
    \addlinespace
    Selecting a materiality benchmark
      & Eval.\ / Evaluative
      & Human leads; LLM provides historical benchmarks \\
    \addlinespace
    Aggregating approved risk indicators
      & Eval.\ / Info
      & Iterative: LLM produces structured analysis; human issues opinion \\
    \addlinespace
    Evaluating compensating-control adequacy
      & Eval.\ / Evaluative
      & Human leads; LLM as evidence aggregator \\
    \addlinespace
    Selecting substantive procedures
      & Anal.\ / Evaluative
      & Iterative: LLM proposes; human selects \\
    \addlinespace
    Audit-evidence extraction
      & Data / Info
      & LLM leads; human reviews for relevance \\
    \addlinespace
    Lead-schedule preparation
      & Data / Info
      & LLM leads with human verification \\
    \addlinespace
    Ratio and trend computation
      & Anal.\ / Info
      & LLM leads; human reviews anomaly flags \\
    \addlinespace
    Audit-memorandum drafting
      & Comm.\ / Evaluative
      & Human leads; LLM supports formatting \\
    \addlinespace
    Audit-report drafting
      & Comm.\ / Evaluative
      & Human leads; LLM supports cross-references \\
    \addlinespace
    Selecting an allowance overlay
      & Anal.\ / Evaluative
      & Iterative: LLM baseline + human overlay \\
    \addlinespace
    Executing an approved PD specification
      & Anal.\ / Info
      & LLM leads; human reviews model adequacy \\
    \bottomrule
  \end{tabular}
  \vspace{4pt}\\
  \footnotesize\textit{Note}: Part 1 of 2; Part 2 continues in
  Table~\ref{tab:matrixdetail2}.
\end{table}

\begin{table}[H]
  \singlespacing
  \small
  \setlength{\tabcolsep}{4pt}
  \renewcommand{\arraystretch}{1.05}
  \centering
  \caption{CDAJ Allocation Detail for 25 Illustrative Work Units
    (Part 2 of 2: Tasks 13--25).}
  \label{tab:matrixdetail2}
  \begin{tabular}{@{}>{\RaggedRight\arraybackslash}p{4.0cm}>{\RaggedRight\arraybackslash}p{3.5cm}>{\RaggedRight\arraybackslash}p{7.5cm}@{}}
    \toprule
    \textbf{Work-unit example} & \textbf{Dominant Stage / Demand} & \textbf{Proposed Allocation} \\
    \midrule
    Significant-increase-in-credit-risk (SICR) determination
      & Eval.\ / Evaluative
      & Human leads; LLM challenges consistency \\
    \addlinespace
    Expected credit loss disclosure drafting
      & Comm.\ / Evaluative
      & Human leads; LLM checks internal consistency \\
    \addlinespace
    Quoted-price fair-value measurement
      & Anal.\ / Info
      & LLM leads with human verification \\
    \addlinespace
    Unobservable-input valuation analysis
      & Anal.\ / Evaluative
      & Iterative: LLM produces candidates; human selects \\
    \addlinespace
    Fair-value option election analysis
      & Eval.\ / Evaluative
      & Human leads; LLM provides scenario fan \\
    \addlinespace
    Revenue recognition policy selection (ASC 606 / IFRS 15)
      & Eval.\ / Evaluative
      & Human leads; LLM supports contract mining \\
    \addlinespace
    Contract-term extraction
      & Data / Info
      & LLM leads; human curates \\
    \addlinespace
    Lease-classification test
      & Anal.\ / Evaluative
      & Iterative: LLM proposes; human signs off \\
    \addlinespace
    Uncertain tax-position assessment
      & Eval.\ / Evaluative
      & Human leads; LLM as adversarial reviewer \\
    \addlinespace
    Populating verified tax-return input fields
      & Data / Info
      & LLM leads with human verification \\
    \addlinespace
    ESG-disclosure data collection
      & Data / Info
      & LLM leads; human reviews for completeness \\
    \addlinespace
    Computing an approved ESG metric
      & Anal.\ / Info
      & LLM leads with human verification \\
    \addlinespace
    Internal-audit report drafting
      & Comm.\ / Evaluative
      & Human leads; LLM supports drafting \\
    \bottomrule
  \end{tabular}
\end{table}

\newpage

\section{Illustrative going-concern record}\label{app:workflow}
This example specifies a record structure, not a finalized case, actual model
output, or observed participant response. In a future case, the evidence file
would contain dated forecasts, debt terms, a default notice, and management's
refinancing representations. The applicable going-concern framework and relevant
assessment horizon must be stated in the case; no generic binary answer can
replace that specification.

\paragraph{Stage A}
Before exposure to AI advice, the professional records an initial assessment,
the evidence supporting it, alternative interpretations, and confidence.
The record distinguishes a documented refinancing commitment from management's
unsupported expectation of obtaining one.

\paragraph{Stage B}
An illustrative recommendation could identify the default notice as evidence
supporting concern, link the claim to its source, and explain why a waiver or
binding financing agreement could change the conclusion. The uncertainty panel
would identify missing confirmation of the financing arrangement. No numerical
confidence is assigned in this mock-up.

\paragraph{Stage C}
The professional records agreement or disagreement, identifies omitted or
misinterpreted evidence, and explains any divergence in reasoning. A response
such as ``management expects refinancing'' requires source evaluation; it is
not automatically stronger evidence than a formal default notice. In C4, a
challenge could ask the model to reconsider the conclusion under the
counterfactual that refinancing is uncommitted. Any reply must be reviewed
against the same evidence file.

\paragraph{Stage D}
The final record preserves the initial estimate, the exact advice and sources,
verification responses, optional exchanges, and the professional's reasoned
conclusion. A change from the initial estimate is neither success nor failure
without reference to the supporting evidence and the independently specified
rubric.

\section{Prospective registration checklist}\label{app:prereg}
Before the proposed human effectiveness evaluation can begin, the research team needs a documented
ethics determination, approved consent and data-processing arrangements, and
feasible recruitment. A separately identified pilot should refine the case
files, scoring ranges, normalization scales, model stimuli, and platform.
Before the confirmatory sample is enrolled, a time-stamped registration should
fix the following:
\begin{enumerate}[leftmargin=2em]
\item The two confirmatory pooled contrasts, outcome definition, multiplicity
procedure, and criteria for a joint conclusion.
\item The four interfaces, common baseline phase, allocation and case-order
procedures, frozen initial advice, version information, and logging fields.
\item Eligibility, fixed recruitment cap and end date, missingness handling,
post-assignment flags, sensitivity analyses, and stopping rules.
\item The panel process, raw reference ranges, independent pilot scales,
scoring code, and evidence that the chosen outcomes are sufficiently informative.
\item The analysis implementation, covariance and distributional assumptions,
convergence fallbacks, and pilot-informed operating-property simulations.
\item The distinction between confirmatory and exploratory analyses, including
case-specific, reliance, time, and confidence-moderation analyses.
\end{enumerate}
Subsequent amendments should preserve the original specification, record dates
and reasons, and distinguish changes informed by outcome data. This checklist
is a proposal, not evidence that any registration has occurred.
\clearpage
\section{Technical details and future evaluation specification}
\label{app:technical}
This appendix combines the technical supplementary material with the preprint.
References to the separate computational archive describe files prepared by the
authors, not a claim of a completed public repository deposit. The new primary
coding analysis is reported separately in Section~\ref{sec:human-coding}; the
original computational archive does not contain private coder responses.
\input{technical_supplement.tex}

\end{document}

%% file: technical_supplement.tex
\subsection{Evidence status}
This supplement documents a public-disclosure retrieval experiment, synthetic
record/event checks, and Gaussian planning computations reported in the main
manuscript. Three real issuer annual reports supply eighteen source-derived
reference cells, not human-subject outcomes or a professionally adjudicated
benchmark. The synthetic inputs and Gaussian draws are separate test data.
No human effectiveness experiment or accounting-model performance benchmark
was conducted. The separate primary coding exercise appears in the main paper;
its private responses are not part of this computational archive. The main
paper also reports all 18 source items from both returns descriptively, including
adverse and unclear statuses. Separate aggregate-only releases preserve the
primary-only and source-descriptive execution deviations; neither is an
expert-gold validation, and the full original human-response plan is unfinished.
The task inventory and formal state/record specification are in the main paper.

\subsection{Aggregate coding contingency tables}\label{supp:primary-matrices}
The following tables reproduce the released aggregate counts. Rows and columns
represent two anonymous response profiles without item identifiers or biography
links. They compare response categories, not predictions against a gold label.
The all-case tables retain explicit U and non-SINGLE parent labels. The two
conditional tables retain only 14 jointly SINGLE cases with nonmissing, non-U
axis codes; U is excluded from these conditional coefficients. Six non-jointly-
SINGLE cases remain visible in the all-case layer. No additional statistic,
post-hoc subset, adjudication, or individual response is introduced.

The original full-plan reader ended in a validation error when numeric realism
answers did not satisfy the categorical schema; no completed full-plan result
is claimed. The separate primary scope was recorded at 08:33:52 UTC before
the first coefficient call at 08:34:40 UTC on 22 November 2026. The separate
source scope was recorded at 09:27:45 UTC before aggregation at 09:28:42 UTC.
The public execution notes preserve these departures without reproducing
private error-output rows. Forty realism answers and the combined fifth
comparison remain unresolved. The original responses are neither recoded nor
replaced by consensus.

\begin{table}[H]\centering\small\singlespacing
\caption{All-case unitization (20 cases).}
\begin{tabular}{lrrrr}\toprule
Profile row / column & SINGLE & SPLIT & UNCERTAIN & Total\\\midrule

SINGLE & 14 & 0 & 0 & 14\\

SPLIT & 1 & 2 & 0 & 3\\

UNCERTAIN & 0 & 0 & 3 & 3\\

\midrule Total & 15 & 2 & 3 & 20\\\bottomrule
\end{tabular}
\end{table}

\begin{table}[H]\centering\small\singlespacing
\caption{All-case stage (20 cases).}
\begin{tabular}{lrrrrrr}\toprule
Profile row / column & S1 & S2 & S3 & S4 & U & Total\\\midrule

S1 & 5 & 0 & 0 & 0 & 0 & 5\\

S2 & 0 & 4 & 0 & 0 & 0 & 4\\

S3 & 0 & 0 & 4 & 0 & 0 & 4\\

S4 & 0 & 0 & 0 & 4 & 0 & 4\\

U & 0 & 0 & 0 & 0 & 3 & 3\\

\midrule Total & 5 & 4 & 4 & 4 & 3 & 20\\\bottomrule
\end{tabular}
\end{table}

\begin{table}[H]\centering\small\singlespacing
\caption{All-case demand response profiles (20 cases).}
\begin{tabular}{lrrrrr}\toprule
Profile row / column & E & I & M & U & Total\\\midrule

E & 7 & 0 & 0 & 0 & 7\\

I & 1 & 4 & 0 & 0 & 5\\

M & 0 & 1 & 4 & 0 & 5\\

U & 0 & 0 & 0 & 3 & 3\\

\midrule Total & 8 & 5 & 4 & 3 & 20\\\bottomrule
\end{tabular}
\end{table}

\begin{table}[H]\centering\small\singlespacing
\caption{Conditional stage (14 comparable cases).}
\begin{tabular}{lrrrrr}\toprule
Profile row / column & S1 & S2 & S3 & S4 & Total\\\midrule

S1 & 5 & 0 & 0 & 0 & 5\\

S2 & 0 & 3 & 0 & 0 & 3\\

S3 & 0 & 0 & 3 & 0 & 3\\

S4 & 0 & 0 & 0 & 3 & 3\\

\midrule Total & 5 & 3 & 3 & 3 & 14\\\bottomrule
\end{tabular}
\end{table}

\begin{table}[H]\centering\small\singlespacing
\caption{Conditional I/E/M demand profiles (14 comparable cases).}
\begin{tabular}{lrrrr}\toprule
Profile row / column & E & I & M & Total\\\midrule

E & 7 & 0 & 0 & 7\\

I & 1 & 4 & 0 & 5\\

M & 0 & 1 & 1 & 2\\

\midrule Total & 8 & 5 & 1 & 14\\\bottomrule
\end{tabular}
\end{table}

\subsection{Archive contents and execution}
The separately prepared reproducibility archive contains \texttt{run\_diagnostics.py},
\texttt{protocol\_harness.py}, \texttt{verify\_results.py}, analytical and Monte
Carlo JSON results, seven compressed raw-draw CSV files, synthetic trace
fixtures with verdicts, and environment information. All three scripts use
only the Python standard library. Extract the archive, run the scripts in this
order, and retain their output:
\begin{verbatim}
python3 run_diagnostics.py
python3 protocol_harness.py
python3 verify_results.py
\end{verbatim}
The scripts locate their own directory, so the shell working directory does
not affect output locations. The first command regenerates simulations and
analytic results; the second regenerates synthetic trace results; the third
checks the raw simulation files. Environment metadata can differ across
computers. Seeded numerical and raw-draw reproducibility was verified in the
recorded Python environment; cross-version pseudorandom implementation changes
should be considered when comparing byte-level hashes.

The output \texttt{legacy\_design\_audit.json} contains auxiliary calculations
illustrating why noninferiority and moderation require different power models.
It is not the current confirmatory family or a human-study registration. Original
manuscript snapshots, original submission materials, and internal reviewer
correspondence are excluded from the public computational archive.

\subsection{Public-disclosure ambiguity diagnostic}
The \texttt{public\_corpus/} subdirectory supplies official URLs, PDF hashes,
limited source cards, acquisition and extraction code, the pre-scoring design
and amendments, retrieval code, derived scores, and verification scripts.
It excludes complete reports, full extracted text, and page images; no open
redistribution license was established. When accessing that archive, consult its README for pinned
\texttt{pypdf==6.21.4} and the exact reacquisition/reconstruction commands.
The main Gaussian and synthetic scripts remain standard-library-only.

The fixed corpus is HSBC Holdings, Barclays, and Standard Chartered annual
reports for 2024: 1,406 complete PDF pages, including three text-empty pages.
Eighteen cards come from six designated ECL and Level-3 tables, selected before
scoring; repeated amounts remain. Card fields identify report, PDF/printed page,
table, row, column, period, unit, and printed value. \texttt{gpt-6-astra-max} assisted
source extraction and visual checks; these labels were not independently
adjudicated by human accountants. Subsequently one human requester confirmed
all 18 source cells and five post-hoc comparisons against an 11-page
source-verification packet, without corrections. The packet hash, scope, and
confirmation record are retained; no checker qualifications or blinding are
claimed, and no frozen label or score changed. The input freeze and source checks are
preserved. A later \texttt{pdfplumber} 0.11.9 cross-check finds all 18 literals
and exact unit strings, with eight multi-occurrence cases retained. Limited
coordinate witnesses and the reconstruction script accompany the public-data
subdirectory; full page-word output stays local. This is different-extractor
corroboration, not semantic table binding or human verification.

The primary comparison is full-field BM25 versus the same BM25 scores with a
value/period/unit token-presence filter. Both use the same value, metric, period,
and unit query; issuer IDs and origin locators are not query hints. Fixed BM25
parameters are $k_1=1.2$ and $b=0.75$. Number-only frequency ranking is a secondary
information-removal ablation. The tokenizer does not reconstruct tables and
strips currency symbols; therefore equal token context does not prove the
correct currency, row, column, or accounting measure.

Both full-field methods recover 13/18 designated origins at rank one and 18/18
within three ranks. The filter changes only one origin's rank from three to
two, increasing mean reciprocal rank from 0.83333 to 0.84259 without any Recall@1 or Recall@3 gain. No origin is excluded and no query abstains. Every card and all 25,308
card-page score rows are retained. The independent checker reconstructs ranks,
ties, gated-score consistency, and aggregates from saved scores. A fresh acquisition of all
three source PDFs reproduced their PDF and per-document extracted-text hashes.
The result is descriptive for eighteen clustered cells, not an inferential
generalization to accounting or evidence of ACP effectiveness.

A post-hoc audit of the five non-origin top-ranked candidates on four unique pages identifies three
repeated balances, one repeated amount with a measure-scope caveat, and one
wrong-measure/column match. The last is Barclays' target Stage~1 mortgage
allowance of \pounds36 million versus a retrieved adjustment \emph{total} of
\pounds36 million whose Stage~1 amount is \pounds7 million. The original
provenance labels and 13/18 result were not changed. This explanatory audit is
not a new semantic-accuracy benchmark. Full methodological limitations and
modest source-locator evidence are in the subdirectory.

\subsection{Finalized-record schema checks}
The \texttt{record\_schema/} subdirectory contains a standalone finalized-record
validator, its field specification, a fabricated loan-portfolio specimen,
33 complete fixtures, and recorded decisions. Execute
\texttt{python3 record\_schema/record\_schema\_validator.py} from the archive
root. Two admissible records and 31 deliberately invalid mutations obtained the
expected verdicts; a rerun reproduced the four JSON outputs. The validator
requires actual initial/final fields, resolvable source and assumption IDs,
matching source-content hashes, tracked contested claims, exactly one documented
exception disposition per listed exception, and a synthetic sign-off.

These are syntactic and reference-integrity checks. They do not establish source
authenticity, semantic support, professional adequacy, genuine signatures,
execution order, or a deployed platform. The source locator is a string, not a
semantic table-cell parser. The specimen's 12,000 arithmetic value is explicitly
not a complete IFRS~9 allowance; its professional conclusion is withheld.
Do not combine the 33 record fixtures and 26 event fixtures into a claim of
integrated system certification.

\subsection{Typed-contract extension (version two)}
The \texttt{record\_schema\_v2/} directory preserves a frozen copy of the original
validator and adds required accounting context and typed source bindings.
Run \texttt{python3 record\_schema\_v2/record\_schema\_validator\_v2.py}.
The accompanying README and source code specify guidance/range status, explicit
currency and unit, document/version/page/table/row/column/measure fields,
printed/normalized values, catalog consistency, and value/period occurrence.
A missing professional basis is marked unresolved or not established, not
filled with fabricated authority or reference bounds.

Forty-two fixtures give 40 structural rejections and two structural acceptances.
One acceptance is the intended fabricated record. The other intentionally
passes despite semantic mismatch: free text asks for a Stage~1 impairment
amount while the jointly changed structured target and binding identify an
adjustment total. This is a negative result about semantic assurance, not a
correct accounting outcome. Changing the supplied binding alone against the
unchanged requested reference is rejected. The supplied catalog is fabricated
and treated as trusted input for consistency checks; it is not independently
established accounting truth. Five JSON outputs reproduce deterministically.
The original 33 record cases, 26 event cases and public retrieval labels/results
remain unchanged; no pooled integrated-system validation claim is made.

\subsection{Statistical algorithm}
Each scenario uses $n=225$ participants per arm and $m=3$ observations. These are
assumed design values, not observed sample sizes. Given correlation $\rho$ and
single-case variance multiplier $v$, independently sample arm means
\[
 \bar Y_2\sim N(0,V),\quad
 \bar Y_3\sim N(d_2,V),\quad
 \bar Y_4\sim N(d_2+d_3,V),\qquad
 V=\frac{v(1+2\rho)}{3n}.
\]
Compute $K2=\bar Y_3-\bar Y_2$ and $K3=\bar Y_4-\bar Y_3$ with known standard
error $\sqrt{2V}$. Their covariance is $-V$ and correlation is $-1/2$.
Two-sided normal $p$-values are adjusted by Holm's two-test rule: the smaller
$p$ must be at most 0.025; only then may the larger be compared with 0.05.
Raw CSV columns record replicate number, both estimates, both $p$-values, and
both rejection indicators.

For each scenario, 100,000 independent repetitions use a dedicated seed.
The seven scenarios are global null; one null and one $-0.40$ increment; both
increments $-0.20$, $-0.30$, or $-0.40$; both $-0.30$ with $\rho=0.60$; and both
raw increments $-0.30$ with variance multiplier 2. Default $\rho=0.30$ and $v=1$.
Seeds are 20261122 through 20261128 in that order. Wilson binomial intervals
quantify Monte Carlo sampling uncertainty. They do not quantify uncertainty in
real-world effect sizes or covariance assumptions.

The main text reports fixed-threshold normal power and MDE separately
from joint Holm rejection. The simulation does not fit a mixed model, estimate
variances, generate actual participant-level behavior, or simulate missingness.
A uniform variance increase at fixed raw effects reduces standardized effects;
it is not a simulation holding standardized $d$ constant.

\subsection{Synthetic validator}
The validator is a reference implementation of selected event-order and
payload-integrity checks. Five acceptable traces and 21 mutated traces form the
26 test fixtures. Inputs and expected/observed decisions are saved in
\texttt{synthetic\_trace\_results.json}. Mutations include missing stages,
condition-inappropriate events, invalid confidence/agreement fields, mismatched
model or frozen-output identifiers, empty verification fields, unpaired or
mismatched refinement messages, and duplicate or premature finalization.

The checks concern encoded fields and ordering, not semantic quality. The
event-fixture schema does not verify the presence or substance of actual initial estimates,
evidence lists, final conclusions, or signatures. It does not establish source
truth, interface visibility, storage immutability, user identity, privacy, or
cross-case isolation. Strict timestamp increase is a test simplification;
production events require a reliable ordering mechanism even when timestamps
tie. A C4 refinement exchange is optional; if begun, its response must be
received before finalization in this reference checker.

\subsection{Independent numerical checks}
The separate verification script rereads all seven raw datasets. It checks
estimated contrast correlation against $-0.5$, sampling variances against the
declared values, and fixed-0.025 rejection rates against analytical normal power.
Tolerance bounds are specified in code. A full rerun reproduced 16 archived
output files in the development environment; the archived internal record
includes inventories that are not scientific outputs and are not shipped as
public source data. The published archive's checksum manifest describes its
actual supplied files.

\subsection{Prospective study versus executed code}
The archive does not contain a deployed participant platform, final evidence
files, adjudicated scoring ranges, pilot scale estimates, actual frozen Stage~B
advice, or a fitted human-outcome analysis. Those would be required to execute
the prospective evaluation. The candidate future primary family is exactly K2 and K3
pooled across cases, subject to the main paper's measurement/feasibility gates; other arm comparisons and all noninferiority or moderation
claims are secondary or exploratory as specified in the main text.

\subsection{AI assistance}
\texttt{gpt-6-astra-max}, an OpenAI model accessed through the configured service,
assisted code preparation, public-source card construction, visual comparison,
and checking. Python execution produced all numerical
values. DeepSeek-v4-pro was used for critique of the revised manuscript and
package, not for generating observations or running the computations. Human
author verification and approval remain necessary before external submission.
\subsection{Candidate human-evaluation agenda: full specification}\label{supp:sec:innovation3}
\subsubsection{Status, scope, and estimands}
No human effectiveness experiment has been conducted. The following is an
evaluation agenda with a candidate contrast structure,
not a finalized confirmatory protocol, registered study, or execution-ready
design. Recruitment agreements, ethics determinations, compensation, the complete
case files, practitioner adjudication, pilot measurements, software deployment,
and frozen model stimuli have not been established. These are prerequisites to
human evaluation, not facts implied by the computational results.

The principal estimands are the intention-to-treat effects of assignment to a
required verification checkpoint (C3 versus C2) and to an interface offering
optional refinement (C4 versus C3), averaged equally over the three specified
cases. They include any extra time, attention, and information induced by the
assigned interface. They do not identify pure cognitive mechanisms, effects of
actual compliance, or general effects across accounting tasks or LLM outputs.

\subsubsection{Cases, participants, and baseline measurement}
The proposed cases cover going concern, IFRS~9 ECL, and Level-3 fair value. Each
requires a bounded evidence file, an explicit reporting/auditing framework and
measurement date, source-linked model advice, and an independent reference
rubric. The three examples in Section~4 of the main text specify work roles;
they are not finalized engagement files. The future sample should consist of
professionals with relevant accounting experience, with credentials and
experience eligibility fixed before screening. Students or CPA candidates
should not be described as credentialed professionals; a student replication
would be a separate population and analysis.

A proposed five-person independent panel, with relevant domain expertise and
no role in writing the cases, would assess case realism and scoring criteria.
For each case, at least 12 prewritten items assessing evidence, assumptions, and justification
would be rated independently on an anchored five-point adequacy scale. The
panel would separately state a defensible numerical range and a written reason
for each boundary. The item-level agreement analysis would report ordinal
Krippendorff's alpha \citep{krippendorff2011} across the item-by-rater matrix, the full disagreement
pattern, and leave-one-rater-out sensitivity. With few items, a point estimate
alone cannot establish reliability. A provisional alpha below 0.80, or material
unresolved disagreement about a scoring boundary, would trigger revision;
the threshold is a proposed design gate, not an observed or universally
sufficient standard. The affected items would be rerated independently after
revision, with original ratings preserved. If disagreement persists, the case
would not enter a confirmatory study. Deliberation must not conceal a genuine
range of defensible interpretations by forcing an artificial consensus.
This procedure is proposed; no panel was recruited and no ratings were obtained.

All participants would complete their initial estimates, evidence lists, and
confidence ratings for all three cases \emph{before random assignment and before
any AI exposure}. This gives genuinely pre-assignment measurements for later
exploratory moderation. Allocation would then be balanced across four arms
within experience strata. A separately randomized case order for the final
phase would be recorded and adjusted for descriptively. Prior exposure within
that phase can still generate learning and carryover; it limits interpretation
of later-case effects. The design estimates a session-level workflow effect,
not isolated independent trial effects.

\subsubsection{Conditions and contrast isolation}
\begin{table}[H]
\centering\singlespacing\small
\caption{Proposed conditions after a common pre-assignment baseline phase.}
\label{supp:tab:contrast}
\begin{tabularx}{\textwidth}{@{}lXXXX@{}}
\toprule
 & C1 & C2 & C3 & C4\\\midrule
Model output & Absent & Frozen output & Same output & Same output\\
Uncertainty panel & Absent & Same qualitative panel & Same panel & Same panel\\
Required verification & No & No & Three questions & Same questions\\
Model challenge interface & Absent & Absent & Absent & Optional refinement\\
Final response & Human & Human & Human & Human\\
\bottomrule
\end{tabularx}
\end{table}
C1 is unaided human judgment; C2 is human judgment with access to a fixed model
recommendation. None is an ``LLM-only'' arm. Common sources, initial estimation,
final-response fields, general task instructions, and the maximum session time
must be identical. C3 adds the Stage~C prompts; C4 additionally offers challenges
and model replies. Extra dialogue is part of C4's treatment, so K3 cannot be
interpreted as iteration independent of additional information or time. There
is no separate AI-alone performance benchmark or evidence of team superiority
over both agents.

Initial Stage~B outputs should be generated once per case, reviewed for study
suitability, frozen, and served identically across AI-assisted arms. The model
identifier, provider version information, prompts, decoding settings, full
outputs, and hashes must be recorded. A model identifier or low temperature
does not guarantee deterministic API behavior. Interactive C4 replies require
complete prompt/response logging; reproducibility means reconstructing the
observed exchanges, not promising identical future generations. No fine-tuning
or accounting-domain adaptation is assumed. Participant information must not
be sent to a third-party API until consent, contractual controls, and
institutional data-governance review permit it.

\subsubsection{Outcome definition and the meaning of accuracy}\label{supp:sec:outcomes}
One candidate primary endpoint is normalized distance from an independently
adjudicated defensible range. For case $j$, let $x_{ij}$ be the participant's
estimate and $[L_j,U_j]$ the reference range. Define
\[
 D_{ij}=\max\{L_j-x_{ij},\ 0,\ x_{ij}-U_j\},\qquad
 Y_{ij}=D_{ij}/s_j,
\]
where $s_j>0$ is a scale fixed from the standard deviation of $D$ in an
independent unaided pilot, before confirmatory outcomes are collected. Lower
$Y$ means closer agreement with the reference range. A zero or unstable pilot
standard deviation requires revising the case or scale before registration;
it cannot be repaired by a data-dependent denominator in the main sample.
Pilot scale uncertainty must be reported and checked using alternative
pre-fixed domain scales.

Going concern would elicit a numerical assessment of substantial doubt with a
separate categorical decision; ECL would elicit the allowance for a specified
exposure or portfolio, not probability of default alone; fair value would elicit
a value in stated currency units. Reference ranges require documented reasons,
including allowable assumptions and treatment of conflicting evidence. Their
width and stability must be inspected for floor effects. These scores are not
inherently on a common 0--100 scale. Normalization makes a pooled contrast
interpretable as an equal-case average in pilot-standard-deviation units; it
does not make professional quality identical across domains.

``Accuracy'' in the propositions is operationalized here as reference
agreement, not objective truth, regulatory compliance, or ex-post prediction
success. A panel can be mistaken or disagree legitimately. Blinded ratings of
evidence completeness and justification quality should be reported separately;
they cannot be silently combined with the distance score. Future case validation
must show that the numerical task and rubric capture the professional judgment
of interest before this endpoint is adopted.

Secondary outcomes include the categorical going-concern decision, active and
total elapsed time, evidence coverage, stated confidence, and checkpoint use.
For exploratory reliance analyses, let $A_{ij}$ and $B_j$ denote the initial
human and frozen model estimates. Report the movement toward advice,
$(x_{ij}-A_{ij})\operatorname{sign}(B_j-A_{ij})$, together with change in
reference distance, $D(x_{ij})-D(A_{ij})$. Zero initial disagreement is reported
separately. Movement toward advice accompanied by increased reference distance
is a descriptive indicator of potentially harmful reliance; it does not by
itself prove an anchoring mechanism. A causal test of advice-error sensitivity
needs randomized advice quality and more stimuli.

\subsubsection{Analysis, multiplicity, and missing data}\label{supp:sec:analysis}
If the measurement gates below support the range-distance endpoint, a
candidate primary analysis would use a repeated-measures model of normalized
reference distance with arm, case, and arm-by-case fixed effects and a
participant random intercept. Estimated marginal means weight the three cases
equally. Restricted maximum likelihood with a small-sample degrees-of-freedom
adjustment is an implementation option to be finalized and tested in the pilot.
Case is fixed because three purposively selected cases cannot support inference
to a population of accounting tasks. The planning calculations in
Section~7 of the main text are not exact power calculations for this fitted
model.

The candidate confirmatory family would contain exactly two two-sided tests: K2=C3$-$C2 and
K3=C4$-$C3, each pooled equally over cases. Holm's procedure \citep{holm1979} controls the family
at 0.05. A negative estimate favors the added element. Report raw and adjusted
$p$-values, estimated contrasts, ordinary 95\% confidence intervals, and
conservative simultaneous 97.5\% intervals for the two comparisons. The ordinary
intervals must not be presented as familywise intervals. Detecting both
increments requires both adjusted tests to reject in the favorable direction;
separate marginal power does not establish power for that joint conclusion.

K1=C2$-$C1, K4=C4$-$C1, all case-specific contrasts, initial-confidence
interactions, reliance outcomes, and time outcomes are secondary or exploratory.
There is no confirmatory noninferiority claim. In particular, a nonsignificant
C4$-$C1 contrast does not establish restored human-level performance. A future
noninferiority study requires a professionally justified margin and its own
sample-size calculation. Confidence is not objective task ambiguity, and one
case per domain cannot identify a general cognitive-demand interaction.

The principal analysis retains all eligible randomized participants in their
assigned arms. Eligibility, consent, and duplicate screening should occur
before randomization. Post-assignment attention failures, speed flags, use of
external assistance, and protocol deviations are reported by arm and used only
in sensitivity analyses, not silently deleted from the intention-to-treat
population. Assignment-to-arm integrity and missing outcomes must also be
reported. Withdrawal and partial completion can prevent outcome observation;
intention-to-treat describes the estimand and does not eliminate missing-data
assumptions.

The model uses available observations under a missing-at-random assumption.
A prespecified imputation sensitivity analysis should include assignment, case,
and pre-assignment estimates, confidence, experience, and familiarity. Session
duration is post-assignment and must not be called a baseline covariate.
Pattern-mixture sensitivity analyses over prespecified plausible shifts should
assess departures from missing at random. Floor effects, heteroscedasticity,
and model convergence must be examined; participant-level resampling and
alternative pre-fixed scales provide robustness checks. Analysis choices and
fallbacks must be finalized using independent pilot data rather than favorable
confirmatory results.

\subsubsection{Recruitment planning and governance}\label{supp:sec:power}
The computations examine a scenario of 225 complete-case-equivalent
participants per arm, or 900 overall. A hypothetical fixed recruitment cap of
1,200 eligible randomized participants, 300 per arm, corresponds arithmetically
to 25\% loss of complete-case information. Neither number is a recruitment
commitment or a recommendation to launch the study. That loss rate is an assumption, not an empirical
estimate. Enrollment would stop at the fixed cap or a predetermined closing
date; it would not continue conditionally on favorable outcomes or realized
per-arm performance. Missingness can reduce information differently across
arms, so 1,200 recruits do not guarantee 225 effective observations per arm.
The pilot must inform feasibility and covariance assumptions before the
confirmatory registration. Any revised cap must be fixed before that sample
is enrolled.

Applicable institutional ethical review or a documented exemption determination
must precede pilot recruitment. Informed consent, fair compensation, withdrawal
rights, pseudonymization, access control, data retention, and third-party model
processing require institution-approved procedures. No ethics approval,
registration or recruitment access is claimed in this paper.
A pilot may refine materials; the final confirmatory protocol, stimuli, scoring
rules, sample cap, analysis code, and amendment policy should then be registered
before the confirmatory field period. A conceptual manuscript and a proposed
registration checklist are not substitutes for that registration.

\subsubsection{Measurement and feasibility gates}\label{supp:sec:gates}
The candidate endpoint deliberately treats values inside the defensible range
as indistinguishable on \emph{reference admissibility}. It cannot be used to
rank their substantive quality. Distance to a panel midpoint would restore
within-range variation but introduce a different assumption---that the midpoint
is more correct than every defensible boundary value. That assumption is not
adopted silently. The independent pilot must therefore compare the range-distance
endpoint with two separately reported sensitivity endpoints: absolute distance
to the panel's defended central estimate, where such an estimate is justified,
and an anchored ordinal rating of evidence use and reasoning. A high floor
proportion, unstable scale, or poor correspondence with independently rated
justification would invalidate range distance as the primary endpoint.

Before choosing the final endpoint, the research team would publish the pilot
score distributions, zero proportions, panel-boundary sensitivity and scale
uncertainty. Pilot-informed simulations must then examine the selected outcome's
distribution, estimated covariance, missingness, and planned model. The present
Gaussian calculations do not clear that gate. Retaining a different primary
endpoint after this pilot is permissible only before the confirmatory
registration; it would require revised power calculations and hypotheses
expressed on that scale.

Frozen advice needs a separate suitability check. Independent reviewers should
code each material advice claim as supported, contestable, or unsupported,
without using later participant outcomes. The pilot must show that the cases
contain an opportunity for the proposed checkpoint to help and a meaningful
possibility that it imposes cost or prompts rejection of useful advice.
If advice quality is not randomized, any estimated K2 or K3 effect remains
conditional on that particular quality mix. No causal claim about sensitivity
to advice errors or a general anchoring mechanism would follow.

A professional minimum important improvement, acceptable added time, target
\emph{joint} power, and recruitment budget must be agreed before sample size is
chosen. Feasibility work should document eligible recruitment rates, completion
burden for both case phases, expected compensation, and access to the intended
professional population. No firm access or budget is assumed. If these gates
cannot support a practicable sample, the team should redesign the evaluation
or report it as a feasibility study rather than call an underpowered study
confirmatory. These open decisions reflect the preliminary status of the human-evaluation
agenda.